\documentclass[manuscript]{acmart}

\AtBeginDocument{%
  }

\setcopyright{acmcopyright}
\copyrightyear{2023}
\acmYear{2023}
\acmDOI{XXXXXXX.XXXXXXX}

\acmJournal{JACM}
\acmVolume{37}
\acmNumber{4}
\acmArticle{111}
\acmMonth{8}

\usepackage{latexsym}
\usepackage{subfig}
\usepackage{arydshln}
\usepackage{booktabs}
\usepackage{multirow}
\usepackage{enumerate}
\usepackage{xcolor}
\usepackage{graphicx}
\usepackage{amsmath}
\usepackage{listings}
\usepackage{xcolor}
\usepackage{float}
\usepackage{xspace}
\definecolor{codegreen}{rgb}{0,0.6,0}
\definecolor{codegray}{rgb}{0.5,0.5,0.5}
\definecolor{codepurple}{rgb}{0.58,0,0.82}
\definecolor{backcolour}{rgb}{0.95,0.95,0.92}
 
\lstdefinestyle{mystyle}{
    commentstyle=\color{codegreen},
    keywordstyle=\color{purple},
    numberstyle=\tiny\color{codegray},
    stringstyle=\color{violet},
    basicstyle=\ttfamily\footnotesize,
    breakatwhitespace=false,         
    breaklines=true,                 
    captionpos=b,                    
    keepspaces=true,                 
    showspaces=false,                
    showstringspaces=false,
    showtabs=false,                  
    tabsize=2,
    frameround=fttt,
    xleftmargin=10pt,
    xrightmargin=10pt,
    language=Java
}

\newcommand{\yun}[1]{\textcolor{black}{{#1}}}

\newcommand{\wcgu}[1]{\textcolor{black}{{#1}}}
\newcommand{\yl}[1]{\textcolor{black}{{{#1}}}}

\newcommand{\revise}[1]{\textcolor{black}{{#1}}}

\newcommand{\tool}{WAVES\xspace}

\begin{document}

\title{Weakly Supervised Vulnerability Localization via Multiple Instance Learning}

\author{Wenchao GU}
\affiliation{%
  \institution{The Chinese University of Hong Kong}
  \city{Hong Kong}
  \country{China}
}
\email{wcgu@cse.cuhk.edu.hk}

\author{Yupan Chen}
\affiliation{%
  \institution{Harbin Institute of Technology, Shenzhen}
    \state{Guangdong}
  \country{China}}
\email{cyp36889@gmail.com}

\author{Yanlin Wang}
\affiliation{%
 \institution{Sun Yat-sen University (Zhuhai)}
 \city{Zhuhai}
 \state{Guangdong}
 \country{China}}
\email{wangylin36@mail.sysu.edu.cn}

\author{Hongyu Zhang}
\affiliation{%
  \institution{Chongqing University}
  \city{Chongqing}
  \country{China}}
\email{hongyujohn@gmail.com}

\author{Cuiyun Gao}
\authornotemark[1]
\affiliation{%
  \institution{Harbin Institute of Technology, Shenzhen}
  \state{Guangdong}
  \country{China}
  \authornote{Corresponding author.}}
\email{gaocuiyun@hit.edu.cn}

\author{Michael R. Lyu}
\affiliation{%
  \institution{The Chinese University of Hong Kong}
  \city{Hong Kong}
  \country{China}}
\email{lyu@cse.cuhk.edu.hk}

\renewcommand{\shortauthors}{Wenchao GU, et al.}

\begin{abstract}
Software vulnerability detection has emerged as a significant concern in the field of software security \yun{recently}, capturing the attention of numerous researchers and developers. 
Most previous approaches 
\yun{focus} on coarse-grained \yun{vulnerability detection,}
such as \yun{at} the function or file level. \yun{However, the developers would still encounter the challenge of manually inspecting a large volume of code inside the vulnerable function to identify the specific vulnerable statements for modification, indicating the importance of vulnerability localization.}\wcgu{
Training the model for vulnerability localization usually requires ground-truth labels at the statement-level, and labeling vulnerable statements demands expert knowledge, which incurs high costs. Hence, the demand for an approach that eliminates the need for additional labeling at the statement-level is on the rise.}
To tackle this problem, we propose a novel approach called \tool for \textbf{W}e\textbf{A}kly supervised 
\textbf{V}ulnerability Localization via multipl\textbf{E} in\textbf{S}tance learning, \wcgu{which does not need the additional statement-level labels during the training.} \tool \yun{has}
the capability to \yun{determine whether a function is vulnerable (i.e., vulnerability detection) and pinpoint the vulnerable statements (i.e., vulnerability localization).}
\wcgu{Specifically, inspired by the concept of multiple instance learning, \tool converts the ground-truth label at the function-level into pseudo labels for individual statements, eliminating the need for additional statement-level labeling. These pseudo labels are utilized to train the classifiers for the function-level representation vectors.} Extensive experimentation on three popular benchmark datasets demonstrates that, in comparison to previous baselines, our approach achieves comparable performance in
vulnerability detection and state-of-the-art performance in statement-level vulnerability localization.
\end{abstract}




\maketitle

\section{Introduction}
\yl{Software vulnerabilities are flaws in the logical design of software or operating systems that can be exploited maliciously by attackers. By exploiting these vulnerabilities, attackers can implant Trojan horses and viruses over networks, extract crucial user information, and even inflict severe damage to the system}\wcgu{\cite{Goodin}}. The detection of software vulnerabilities has emerged as a crucial issue in the realm of software security, garnering considerable interest from researchers and developers in recent decades.

Most traditional vulnerability detection tools ~\cite{Checkmarx, Flawfinder, Infer, ITS4, SVF}, such as Flawfinder{~\cite{Flawfinder}}, employ static analysis techniques to identify vulnerabilities in programs. These tools typically rely on predefined vulnerability patterns to determine whether the target programs are vulnerable. Although these tools can effectively detect well-defined vulnerabilities such as use-after-free issues, they often struggle to identify vulnerabilities that are not easily defined. Furthermore, the manual definition of vulnerability patterns is a time-consuming process that can hinder the efficiency of these methods. Moreover, these tools often generate a large number of false positives/negatives in their reported vulnerabilities~\cite{DeepWukong}, further diminishing their utility.

With the advancement of deep learning techniques, there has been a growing interest in using these methods for vulnerability detection in recent years. Various deep learning-based approaches have been proposed by researchers, leveraging neural networks like Convolutional Neural Networks (CNNs)~\cite{Russell}, Recurrent Neural Networks (RNNs))~\cite{LSTM}, and Graph Neural Networks (GNNs)~\cite{devign, reveal}. These approaches enable automatic acquisition of vulnerability features or patterns from training data. Notably, these techniques have demonstrated their effectiveness in detecting unreported or unknown vulnerabilities~\cite{sysevr}.

However, most current deep learning-based approaches for vulnerability detection only offer predictions at the function level. This falls short of developers' needs because the majority of vulnerable statements within the code tend to be relatively concealed and challenging to uncover. Even with function-level predictions, developers still face the time-consuming task of locating these vulnerable statements. Some previous approaches~\cite{linevul,IVDETECT} have aimed to enhance the interpretability of models to help developers identify vulnerable statements. However, these methods~\cite{linevul,IVDETECT} have struggled to achieve accurate localization, as they do not prioritize the localization problem during training and solely rely on attention scores or GNN explainers
~\cite{GNNExplainer} to explain the model's behavior after training. Automatically predicting statement-level vulnerabilities in a supervised manner poses difficulties, as it necessitates labeled data for model learning. Therefore, there is an urgent demand for unsupervised or weakly supervised approaches for statement-level \wcgu{vulnerability localization}.

Multiple instance learning (MIL) is a form of weakly supervised learning that finds extensive use across diverse tasks, including drug activity prediction~\cite{BergeronMZBB12}, image retrieval~\cite{RahmaniG06,AndrewsTH02}, and text classification~\cite{DBLP:conf/kdd/KotziasDFS15}. MIL handles training data arranged in sets called bags, where each bag contains multiple instances. In MIL, only the label for the entire bag is provided, while the labels for individual instances are unknown. MIL allows us to convert the bag's label into pseudo-labels for each instance, thus addressing the missing label problem during instance classification training. This scenario is similar to vulnerability localization, where only the vulnerability label for the entire function is given, and the labels for each statement are unavailable. And we discovered that we can effectively reframe the problem of vulnerability localization as a Multiple Instance Learning (MIL) problem. In this approach, each statement within the target function is treated as an instance, and the entire function represents a bag. Since each bag (fucntion) contains multiple instances (statements), it follows that a vulnerable function must include at least one vulnerable statement. Conversely, if a function is not vulnerable, all its included statements are considered non-vulnerable. This logic parallels the relationship between the label of a bag (function) and the labels of its instances (statements) in MIL. However, applying MIL to this task faces two primary challenges. Firstly, when employing deep learning models for code analysis, the smallest unit of input is typically variable names or subtokens, rather than whole statements. This complicates the generation of statement-level representation vectors that seamlessly integrate with the MIL framework. Secondly, traditional MIL approaches assume independence among instances within the same bag, implying no interaction across different instances~\cite{CarbonneauCGG18}. However, in vulnerability detection, a single statement often does not raise vulnerability concerns; vulnerabilities typically arise from specific contextual statements within the program. To tackle these challenges, we propose a Transformer-based model within the framework of multiple instance learning. 
This model can effectively capture local and global vulnerability information for each statement and allows statements within the same function to interact during the training. Additionally, it retains the core concept of generating pseudo instance labels from the bag label. By adopting this approach, we can construct pseudo-labeled training instances, reducing the labor-intensive task of manually labeling vulnerabilities at the statement level.

In this paper, we propose a novel approach named \tool for function-level vulnerability detection with statement-level localization. \tool first converts an input code snippet into a token sequence and feeds it into a Transformer-based encoder. During the encoding process, tokens from the same or different statements interact freely, enabling the model to learn contextual information for each statement. There are two channels aiming to capture local and global features separately. The statement-level classifier for each channel is then trained individually to determine whether the statement-level representation vectors are vulnerable or not. The results from these two classifiers are combined to produce a single prediction for a single statement. The evaluation of \tool is conducted using three widely used datasets, and extensive experimental findings showcase that \tool achieves comparable performance in detecting vulnerabilities at the function level compared to previous models. Furthermore, its ability to localize vulnerabilities surpasses that of the previous models.

We summarize the main contributions of this paper as follows:

\begin{itemize}

\item We introduce \tool, a novel approach that achieves vulnerability localization without requiring additional statement-level labels during model training. Furthermore, the statement-level predictions from \tool can be leveraged for function-level vulnerability detection. To our best knowledge, \tool is the first approach to adopt multiple instance learning for localizing vulnerabilities at the statement level, all without requiring additional vulnerability labeling at the statement level.

\item We integrate various pooling modules capable of capturing code features specific to vulnerabilities. We also validate the effectiveness of each pooling module on the overall performance.

\item We have performed comprehensive experiments on public benchmarks, and the results indicate that \tool achieves comparable performance in function-level vulnerability detection and outperforms previous models in statement-level vulnerability localization, showcasing state-of-the-art performance.
\end{itemize}

The remainder of this paper is structured as follows: Section~\ref{sec:method} provides an overview of the architecture of our proposed \tool, along with its design details. Section~\ref{sec:setup} describes our experimental setup, including the datasets used, evaluation metrics, and implementation specifics. In Section~\ref{sec:results}, we present the experimental results and provide our analysis. In Section~\ref{sec:threats}, we discuss the threats to the validity of our experiments. \wcgu{Section~\ref{sec:suvery} discusses the related work on vulnerability detection and multiple instance learning, while Section~\ref{sec:conclusion} concludes the paper. }

\section{Methodology}
\label{sec:method}

In this section, we propose a novel deep learning-based vulnerability detection approach that can achieve function-level detection and statement-level localization simultaneously with weakly supervised learning. Specifically, we present the overview and detailed design of the proposed approach \tool, \yl{including
model design, model training strategy, and inference strategy}.

\subsection{Overview}
\begin{figure*}[ht]
\centering
\includegraphics[width=1\textwidth]{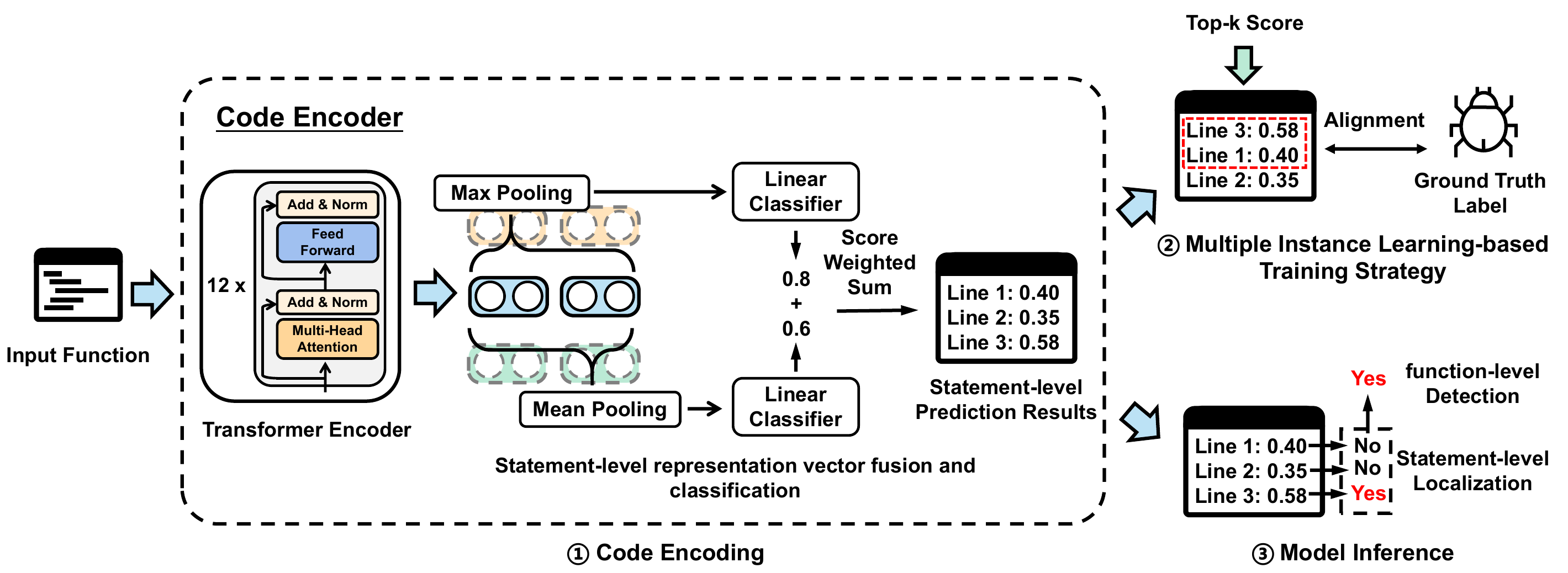}
\caption{An overview architecture of \tool, containing three main steps. \textcircled 1 Code Encoding: The target function will first be transformed into a token sequence and then fed into a Transformer-based Encoder. Subsequently, the token vectors will be integrated into statement-level vectors, and two linear classifiers will be employed to classify them. \textcircled 2 Multiple Instance Learning-based Training Strategy: The statements will be ranked in descending order based on the statement-level predicted results.
The top-k statements will then be assigned pseudo labels identical to the function label for model training purposes. \textcircled 3 Model Inference: The results obtained in Step 1 will be utilized to predict the vulnerability of each statement. The prediction results from all the statements will then be used to determine the vulnerability of the entire function.}
\label{fig:framework}
\end{figure*}

\wcgu{Figure~\ref{fig:framework} illustrates an overview of the proposed approach \tool. Our approach consists of three steps: code encoding, multiple instance learning-based training strategy, and model inference. In the step of code encoding, the Transformer-based encoder learns to generate representation vectors for each statement within the given function. Two linear classifiers are trained to classify whether these statement-level representation vectors indicate vulnerability or not. In the multiple instance learning-based training strategy, we convert function-level ground-truth labels for vulnerability detection into statement-level pseudo labels and utilize these pseudo labels for the model training. During the inference step, the model also generates representation vectors for each statement and employs the previous two trained linear classifiers to determine their vulnerability status. Additionally, the overall vulnerability prediction for the entire function is determined by considering the vulnerability predictions of each statement within the function. Further details about these three steps will be presented in the following sections.}

\subsection{Code Encoding}

Following previous work~\cite{linevul}, we adopt the Transformer as the base model for code encoding. The given code snippet will be converted into a token sequence and each token will be split into sub words by the tokenizer. We adopt the Byte Pair Encoding (BPE) approach~\cite{BPE} from Roberta as our tokenizer to tokenize the word. To incorporate the positional relationships between the subwords into the model, positional embedding vectors are encoded to represent the token's position within the token sequence. The token embedding vector and the positional embedding vector are then combined into a unified representation vector, which represents the corresponding token within the input sequence.

It is necessary to record the location information of each token for the generation of statement-level representation vectors in subsequent stages. In order to capture this information, we create a binary statement indicative matrix. The matrix, denoted as $S$, is defined as follows:

\begin{equation}
S=\{s_{11},...,s_{1n},...,s_{m1},...,s_{mn}\}
\end{equation}

\noindent where $n$ is the token number, $m$ is the statement number, and $s_{ij}$ indicates whether the i-th token belongs to the j-th statement in the given function. The value of $s_{ij}$ will be 1 if the i-th token belongs to the j-th statement in the given function; otherwise, $s_{ij}$ will be zero. 

\wcgu{In the subsequent sections, we will introduce the design of the code encoder, which takes both the token embedding sequence and the binary statement indicative matrix as inputs and generates statement-level prediction scores.}

\subsection{The Design of Code Encoder}
\label{sec:model}
In this subsection, we present the \wcgu{code encoder} of our proposed approach \tool. The \wcgu{code encoder} consists of a Transformer-based encoder, as well as linear classifiers for both the max pooling channel and the mean pooling channel.

\subsubsection{Transformer Encoder with Self-attention}
In our approach \tool, we utilize a Transformer-based encoder. The encoder comprises 12 stacked Transformer blocks, each consisting of a multi-head self-attention layer and a fully-connected feed-forward neural network. The multi-head self-attention layer's purpose is to generate the attention vector based on the attention score assigned to each code token. To accomplish this, the dot product between the query vector of the current code token and the key vectors of the other tokens is computed. Subsequently, the dot product is normalized to probabilities via the Softmax function. Finally, the attention vector is obtained by taking dot product between the value vectors and previous normalized probabilities. The equation for calculating the attention score is provided below:
\begin{equation}
    Attention(Q,K,V) = {\rm Softmax} (\frac{QK^T}{\sqrt{d_k}})V,
\end{equation}

\noindent where $Q$, $K$ , $V$ is the query vector, key vector, and value vector, respectively.

The multi-head mechanism enables the model to create several subspaces, each dedicated to different aspects of the input sequence. This allows the model to effectively capture diverse semantic information from the input. Initially, the multi-head mechanism divides the input vectors into $h$ heads, with each head having a dimension of $\frac{d}{h}$. Following the self-attention operation on each head, these heads are then concatenated back together as:

\begin{equation}
\begin{aligned}
MultiHead & (Q,K,V)  = Concat(head_1,...,head_h)W^O,
\end{aligned}
\end{equation}

\noindent where $head_i = Attention(QW^Q_i,KW^K_i,VW^V_i)$ and $W^O$ is the projection matrix for the concatenated vectors.

Finally, the concatenated vectors will be fed into a fully-connected feed-forward neural network. This neural network comprises two linear layers, with a ReLU activation function sandwiched between them. As demonstrated by numerous previous studies~\cite{codebert,GuoRLFT0ZDSFTDC21,GuoLDW0022}, Transformers have advantages over RNNs in capturing long-term dependencies. They are more effective at capturing contextual information among different statements, even with variations in edit distance. This is why we chose Transformers for encoding the code.

\subsubsection{Statement-level Representation Vector Fusion}
After the encoding of the input token sequence from the previous Transformer-based encoder, we obtained the representation vector for each token from the last hidden states in the model. Since previous Transformer-based approaches focus solely on function-level vulnerability detection, they do not require generating statement-level representation vectors. In contrast, our approach~\cite{linevul} aims to achieve statement-level vulnerability localization, making it crucial to generate these vectors. To accomplish this, we need to merge the embedding vectors of tokens within the same statement into a single statement-level representation vector. We propose two methods for efficiently fusing these vectors: max pooling and mean pooling. The max pooling method captures local suspicious information within a statement, while the mean pooling method captures suspicious information from a broader, statement-wide perspective.

\noindent \textbf{Max Pooling Channel}: Max pooling is a down-sampling technique frequently used in deep learning to efficiently preserve local features from the original data. This property enables the model to capture local information, such as variable or API misuse, which is crucial for software vulnerability detection. The operation of max pooling is illustrated as follows:
\begin{equation}
    v_{max\_j} = max(h_{1} \cdot s_{1j},...,h_{n} \cdot s_{nj})
\end{equation}
\noindent where $v_{max\_j}$ is the representation vector for the locality information in the j-th statement, $h_{i}$ is the hidden vector for the i-th token from the encoder, and $s_{ij}$ is the indicator to show whether the i-th token belongs to the j-th statement. $h_{i} \cdot s_{ij}$ can remove the irrelevant token vectors during the operation of max pooling.

\noindent \textbf{ Mean Pooling Channel}: Mean pooling is a down-sampling technique frequently used in deep learning to efficiently preserve global features from the original input. This technique assists the model in determining if the execution of a single statement might raise a vulnerability issue. The operation of mean pooling is illustrated as follows:
\begin{equation}
    v_{mean\_j} = \frac{\sum_{i=1}^{n} h_{i} \cdot s_{ij}}{\sum_{i=1}^{n} s_{ij}}
\end{equation}
\noindent where $v_{mean\_j}$ is the representation vector for the global information in the i-th statement,  $h_{i}$ is the hidden vector for the i-th token from the encoder, and $s_{ij}$ is the indicator to show whether the i-th token belongs to the j-th statement. $h_{i} \cdot s_{ij}$ can remove the irrelevant token vectors during the operation of mean pooling.

\lstset{language=C}
\begin{lstlisting}[caption={A motivating example for mean pooling.},label={lst:mean_example}]
#define JAN 1
#define FEB 2
#define MAR 3

short getMonthlySales(int month) {...}

float calculateRevenueForQuarter(short quarterSold) {...}

int determineFirstQuarterRevenue() {

// Variable for sales revenue for the quarter
float quarterRevenue = 0.0f;

short JanSold = getMonthlySales(JAN); /* Get sales in January */
short FebSold = getMonthlySales(FEB); /* Get sales in February */
short MarSold = getMonthlySales(MAR); /* Get sales in March */

// Calculate quarterly total
short quarterSold = JanSold + FebSold + MarSold;

// Calculate the total revenue for the quarter
quarterRevenue = calculateRevenueForQuarter(quarterSold);

saveFirstQuarterRevenue(quarterRevenue);

return 0;
}
\end{lstlisting}

Here is a motivating example from CWE-190, specifically "Integer Overflow" or "Wraparound," presented in Listing~\ref{lst:mean_example}. In line 19, the values of the variables \texttt{JanSold}, \texttt{FebSold}, and \texttt{MarSold} are summed and assigned to the variable \texttt{quarterSold}. However, there is a potential risk of integer overflow if the sum exceeds the maximum value allowed for the \texttt{short} int data type. In this example, the max pooling channel can help the model focus on the data type of \texttt{quarterSold} being \texttt{short}, and the mean pooling channel can help the model notice the sum operation in this statement. By combining the information from these two channels, the model can detect a potential data overflow problem.

\subsubsection{Representation Vector Classification and Fusion}
Upon combining the original embedding vectors of the tokens within a statement, we generate two representation vectors that encompass both local and global information. Then we employ two linear classifiers to classify these two representation vectors as the binary classification task, respectively. Each classifier consists of a fully connected layer followed by a softmax activation function. The scores produced by these classifiers are then combined through a weighted sum, resulting in a single score that serves as the final prediction for the statement. This fusion enables the model to assess whether a given statement is vulnerable by considering both local and global perspectives, thereby enhancing the model's detection capability.

\begin{figure*}[ht]
\centering
\includegraphics[width=0.9\textwidth]{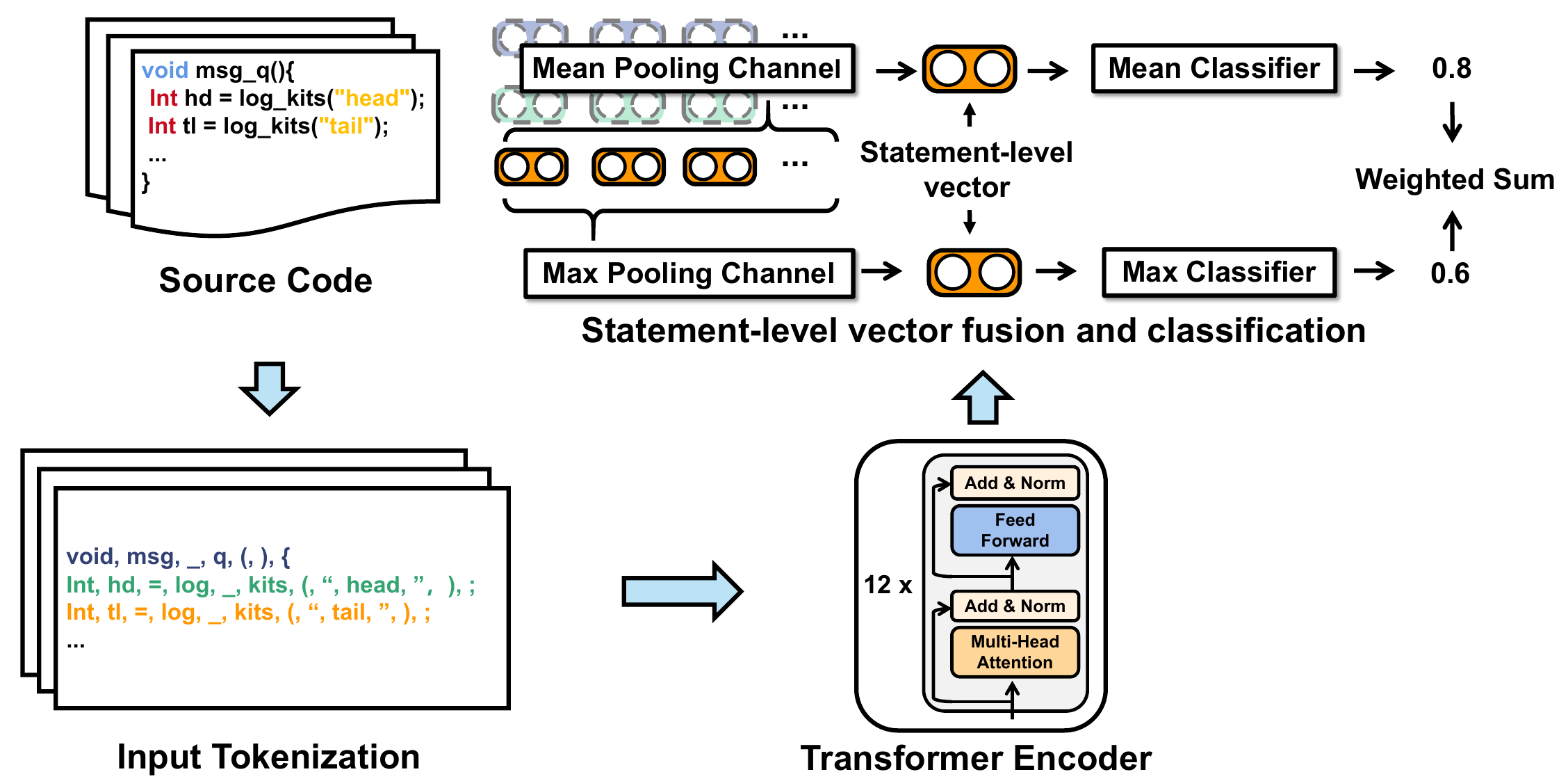}
\caption{An illustrative example of code encoder design}
\label{fig:example}
\end{figure*}

Figure~\ref{fig:example} illustrates our code encoder design. Initially, the code is split into several statements, and each statement is tokenized into a sequence of tokens. These token sequences are then fed into a Transformer-based encoder, where each token is encoded into a vector. Next, the token vectors belonging to the same statement are fused into two statement-level vectors via mean pooling and max pooling channels, respectively. Finally, the statement vectors are fed into two classifiers, one for mean pooling and one for max pooling. Predictions from the local and global views are obtained and combined into a single prediction score.

\subsection{Multiple Instance Learning-Based Training Strategy}
Despite generating the representation vector and utilizing the classifier for individual statements, the absence of statement-level labels poses an ongoing challenge. Inspired by the concept from multiple instance learning, we convert the label of the entire function into pseudo labels for each statement within the target function to tackle this issue. These pseudo-labels serve as the supervised signal during training.

It is known that a vulnerable function must contain at least one vulnerable statement, whereas a non-vulnerable function has no vulnerable statements. In a given code snippet, the label $y_i$ represents the vulnerability status of i-th statement, where $y_i$ is 0 for non-vulnerable statements and 1 for vulnerable statements. The label $Y$ indicates whether the code snippet as a whole is vulnerable or not. Thus, we can determine the label of the entire code snippet as follows:

\begin{equation}
    Y={\rm max}\{y_1,...,y_n\}
\label{eq:overall}
\end{equation}

\noindent where $n$ is the number of statements in the code snippet.

By referring to equation~\ref{eq:overall}, it becomes evident that the label of every statement within a function labeled as 0 will also be 0. However, there are two challenges in determining the statement labels within a function labeled as 1. Firstly, the vulnerable function contains only a few vulnerable statements, while the majority of statements inside the function are non-vulnerable. This raises the issue of how to assign labels to these statements. Secondly, even if we solve the problem of label assignment for the vulnerable function, we encounter another challenge with the label ratio. In non-vulnerable functions, the statement labels will be predominantly 0, and the same applies to most of the statement labels in the vulnerable function. The ratio of pseudo-positive/negative labels becomes much smaller than the ratio of the original positive/negative labels in the dataset. \wcgu{The sample ratio imbalance will lead the model to favor predicting functions as non-vulnerable, impacting its overall performance~\cite{LinGGHD20}.}

We address the first issue based on the assumption that the non-vulnerable statements exhibit a distinct pattern that differs from the vulnerable statement. This pattern can be detected and distinguished by the classifier model. Specifically, we sort the statements within the function in descending order, based on their previous classification prediction scores. Next, we generate pseudo labels for the top k statements, assigning them the same label as the entire function for training purposes. The value of k can be determined as the average number of vulnerable statements, as indicated by the dataset statistics. It is important to note that the pseudo labels for the vulnerable statements are not accurate at the beginning of the training. However, the pseudo labels for the non-vulnerable statements must be accurate because all the statements in the non-vulnerable statements are non-vulnerable. As negative samples are incorporated into the training process, the prediction scores for statements that are semantically similar to non-vulnerable statements will progressively decrease. The pseudo labels for vulnerable statements will gradually become more relatively accurate as training progresses~\cite{CarbonneauCGG18}.

To address the second issue, we set the training objective exclusively for the top-k statement within the given function, regardless of whether the function is vulnerable or non-vulnerable. \wcgu{This operation can effectively address the issue of imbalanced sample ratios resulting from the conversion of pseudo labels.} 
From the experiment results, we find that the selection of hyperparameter k will have a slight effect on the overall performance, which will be discussed in Section~\ref{sec:rq2}

In summary, the loss function used in \tool is cross-entropy loss, which is defined as follows:

\begin{equation}
\begin{aligned}
    loss=\frac{1}{N \cdot k}\sum_i \sum_j^k -[Y_i log(p_{ij}) + (1-Y_i) log(1- p_{ij})]
\end{aligned}
\end{equation}

\noindent where $N$ is the number of the function, $Y_i$ is the label for the function $i$, $p_{ij}$ is the vulnerability probability for the j-th statement in the function i, and k is the pre-defined parameter.

\subsection{Model Inference}
\label{sec:infer}
During the inference stage, \tool does not provide a direct prediction of the vulnerability of the entire function. Instead, \tool focuses on predicting vulnerable conditions for each statement within the function. The determination of the function-level prediction relies on the results obtained at the statement level. If at least one statement within the function is predicted as vulnerable, the function is considered vulnerable as well. Conversely, if no statements are predicted as vulnerable, the function is deemed non-vulnerable.

\wcgu{Two methods for localizing vulnerabilities are employed: absolute label prediction and relative scores ranking. In the absolute label prediction method, the model identifies and reports only the statements it predicts as vulnerable. On the other hand, the relative scores ranking method involves evaluating the statements within the function based on their prediction scores and sorting them in descending order. The top k statements are then selected as potential candidates for vulnerable statements, which are presented to the users. However, it is important to note that the absolute label prediction might not be accurate enough since there is no ground-truth label provided for training. Therefore, we recommend utilizing the relative scores ranking method as the preferred approach for vulnerability localization. The performance of both methods will be further discussed in the following section.}
\section{Experimental Setup}
\label{sec:setup}

In this section, we provide an overview of the statistics information for the dataset used in our study, the steps taken for data pre-processing, the baseline models employed, the evaluation metrics utilized, and the implementation details concerning both our proposed tool and the other baseline models included in our experiment.

\begin{table}[ht]
\small
\setlength\tabcolsep{5pt}
\centering
\caption{Statistics of dataset.}
\begin{tabular}{llllll}
\toprule
\multicolumn{2}{c}{\textbf{Dataset}} & \textbf{Vul function} & \textbf{Non-vul function} & \textbf{Avg stat num} & \textbf{Avg vul stat num}\\
\midrule
\multirow{3}{*}{\textbf{Fan \textit{et al.}}} & \textbf{Training} & 4,993 & 142,188 & 20.12 & 3.03 \\
& \textbf{Validation} & 624 & 17,774 & 20.69 & 3.27 \\
& \textbf{Testing} & 626 & 17,774 & 20.41 & 3.28 \\
\midrule
\multirow{3}{*}{\textbf{CVEfixes}} & \textbf{Training} & 4,437 & 7,699 & 72.14 & 7.69 \\
& \textbf{Validation} & 554 & 964 & 73.14 & 7.60 \\
& \textbf{Testing} & 588 & 960 & 71.79 & 5.97 \\
\midrule
\multirow{3}{*}{\textbf{Reveal}} & \textbf{Training} & 801 & 10,371  & N/A & N/A \\
& \textbf{Validation} & 98 & 1,256 & N/A & N/A\\
& \textbf{Testing} & 104 & 1,296 & N/A & N/A \\
\midrule
\multirow{3}{*}{\textbf{FFMPeg+Qemu}} & \textbf{Training} & 7,078 & 8,526 & N/A & N/A \\
& \textbf{Validation} & 887 & 1,058 & N/A & N/A \\ 
& \textbf{Testing} & 879 & 1,024 & N/A & N/A\\ 
\bottomrule
\end{tabular}
\label{tab:stat}
\end{table}

\subsection{Data Pre-processing}
In our experiment, we evaluate the performance of vulnerability detection and localization using four datasets: FFMPeg+Qemu~\cite{devign}, Reveal\cite{reveal}, Fan et al.~\cite{fan}, and CVEfixes~\cite{BhandariNM21}. The FFMPeg+Qemu dataset, collected by Devign, comprises data from two open-source C projects and has been labeled by experts. It consists of approximately 10,000 vulnerable functions and 12,000 non-vulnerable functions. The Reveal dataset is obtained from Linux Debian Kernel and Chromium, containing about 2,000 vulnerable functions and 20,000 non-vulnerable functions. Fan et al. dataset, a C/C++ dataset, is gathered from over 300 open-source GitHub projects, covering 91 different Common Vulnerabilities and Exposures (CVE) databases from 2002 to 2019. This dataset includes around 10,000 vulnerable functions and 177,000 non-vulnerable functions. \revise{The CVEfixes dataset compiles CVE records from the National Vulnerability Database (NVD). Its initial release covers 5,365 CVEs from 1,754 projects as of June 9, 2021. For our experiments, we extract C/C++-related CVEs from this dataset.}

The FFMPeg+Qemu and Reveal datasets only provide labels indicating whether a given function is vulnerable or not. Therefore, we solely assess the performance of function-level vulnerability detection using these two datasets. In contrast, Fan et al.~\cite{fan} and CVEfixes~\cite{BhandariNM21} not only offer function-level labels but also provide the fixed version of the vulnerable function. This allows us to pinpoint the vulnerable statements by comparing the function before and after fixing. Therefore, we can evaluate both function-level vulnerability detection and statement-level vulnerability localization on this dataset.

We have imposed a length limitation on the input token sequence in order to accommodate the fixed-length input requirement of Transformer. Any token exceeding the maximum input length is discarded. However, the vulnerable statements in the code snippets may be contained in the discarded tokens, which means that the input statements are not vulnerable although the label for the entire function is vulnerable. To address this label conflict, we remove code snippets whose function-level label is vulnerable, but do not contain any vulnerable statements in the input to our model. Additionally, different baseline models used in our experiments necessitate different data pre-processing tools. Some of the data in our dataset cannot be processed correctly by all of these tools. In the interest of experimental fairness, we only retain the data that can be processed by all data pre-processing tools. To maximize the length of code that can be processed by the model, we removed all comments, ensuring that the input consists solely of executable code.

Within Table~\ref{tab:stat}, "Vul function" denotes the number of vulnerable functions in the dataset, while "Non-vul function" represents the number of non-vulnerable functions. "Avg stat num" indicates the average number of statements in a single function within the dataset, and "Avg vul stat num" signifies the average number of vulnerable statements in a single function. As previously mentioned, Reveal and FFMPeg+Qemu do not provide information regarding vulnerable statements, thus preventing us from offering statistics on the average statement number and average vulnerable statement number for these two datasets.

\subsection{Implementation Details}
 In our proposed \tool, we set the number of encoder layers to 12, the number of attention headers to 12, and the hidden size to 768. The batch size and learning rate were set to 16 and 2e-5, respectively. Our model supports a maximum input token length of 512. As the hyperparameter top-k is sensitive to the average number of vulnerable statements in the target dataset, which can vary between datasets, we experimented with values of 1, 3, and 5 for top-k and selected the best performance for each dataset. For optimization, we utilized the AdamW~\cite{adamw} optimizer. We set a maximum of 50 epochs for the training with 10-steps patience for early stopping. 

We replicated all the baselines, except for Devign, using publicly released source code and adopted the same hyperparameter settings as described in their original paper. For Devign, since they did not make their code public, we reproduced it based on the code provided by ~\citet{reveal}. In the case of the baseline model called IVDetect, it requires the training dataset to have an equal ratio of vulnerable and non-vulnerable functions. Since none of the datasets we used had this ratio, we retained all the vulnerable functions and randomly selected an equal number of non-vulnerable functions from each training dataset to create a new training dataset for IVDetect. There are two version of the baseline model named LineVul, which are LineVul with pre-training and LineVul without pre-training. We observed that LineVul with pre-training performed exceptionally well on the dataset of Fan et al, while there was no significant difference between LineVul with and without pre-training on the other two datasets. This led us to suspect the presence of a data leakage problem specifically in the Fan et al dataset so that we exclude the pre-training model from our experiments. All the models were trained on a server equipped with NVIDIA A100-SXM4. The training process for \tool consumed approximately 10 GPU hours.

\subsection{Baselines}
We compare \tool with six state-of-the-art vulnerability detection methods, including two token-based methods~\cite{vuldeepecker, sysevr}, three structure-based methods~\cite{devign, reveal, IVDETECT}, and one unsupervised statement-level detection method~\cite{linevul}. Here, we provide a brief description of these baseline methods:

\textbf{(1) VulDeePecker} \cite{vuldeepecker}: VulDeePecker extracts code gadgets from the given code snippet, which are several lines of code that are semantically related to each other, and adopted the bidirectional LSTM-based neural network with an attention mechanism for vulnerability detection.

\textbf{(2) SySeVR} \cite{sysevr}: SySeVR extracts the vulnerability syntax characteristics (SyVCs) from the given function at first and then transforms these SyVCs into semantics-based vulnerability candidates (SeVCs) which contain the statements related to the given SyVCs via the program slicing technique. Finally, a bidirectional recurrent neural network is employed to encode these SeVCs into vectors, facilitating the detection of vulnerable code snippets.

\textbf{(3) Devign} \cite{devign}: Devign extracts the information of abstract syntax tree (AST), control flow graph (CFG), data flow graph (DFG), and code token sequence from the given function to construct the graph which can represent the given functions and generate the embedding vector for each node inside the graph. Subsequently, the graph is inputted into a Gated Graph Neural Network (GGNN) for classification training.

\textbf{(4) Reveal} \cite{reveal}: Reveal extracts the information of Code Property Graph (CPG) from the given function to construct the representation graph and adopts the technique of Word2Vec to generate the embedding vector for each node. The resulting graph is then inputted into GGNN, where all the vectors from the representation graph are combined into a single vector. This fused vector serves as the representation for the entire graph, facilitating vulnerability detection.

\textbf{(5) IVDetect} \cite{IVDETECT}: IVDetect is a code analysis tool that divides the code into multiple statements and extracts various features from each statement. These features encompass sub-token sequences, AST sub-trees, variable names, variable types, data dependency context, and control dependency context. To capture these features, they are embedded into representation vectors. Subsequently, these vectors are combined into a unified representation vector for each statement using an attention-based bidirectional GRU. These statement-level representation vectors serve as node embedding features, which are then fed into a Graph Convolutional Network (GCN) to acquire a comprehensive graph representation for detection purposes.

\textbf{(6) LineVul} \cite{linevul}: LineVul adopts the Byte Pair Encoder (BPE) technique to tokenize the given code into a sub-token sequence and utilize a 12-layer Transformer based model for vulnerability detection. Not only predicting the function-level vulnerability, LineVul can also localize the vulnerable statement by calculating the attention scores for each sub-token.

\subsection{Evaluation Metrics}
In our experiment, we adopts four metrics, which are $ACC$, $P$, $R$, and $F1$, to evaluate the performance of all the models in function-level vulnerability detection. Additionally, we utilized nine metric, which are $Top-1$, $Top-5$, $Top-10$, $MFR$, $MAR$, $IFA$, $P$, $R$, $F1$, to evaluate the performance of all the models in statement-level vulnerability localization.

The metric employed to determine the accuracy of the model is denoted as $Acc$. It quantifies the ratio of accurately classified samples to the total number of samples. The definition of $Acc$ is presented below:

\begin{equation}
    Acc = \frac{S_c}{S}
\end{equation}

\noindent where $S_c$ represents the number of samples correctly labeled by the model, while $S$ denotes the total number of samples. A higher value of $Acc$ indicates a better performance of the model.

$P$ is the metric used to assess the accuracy of the model's detection of vulnerable samples. The definition of $P$ is provided below:

\begin{equation}
    P = \frac{TP}{TP+FP}
\end{equation}

\noindent where $TP$ represents the count of samples where both the label and the model's prediction are true, while $FP$ refers to the count of samples where the label is true but the model's prediction is incorrect. A higher value for $P$ signifies improved performance of the model.

$R$ is a metric used to assess the percentage of vulnerable samples correctly detected out of all the vulnerable samples predicted by the model. The specific definition of this metric is provided below:

\begin{equation}
    R = \frac{TP}{TP+FN}
\end{equation}

\noindent where $TP$ represents the count of samples with true labels that the model correctly predicts, while $FN$ represents the count of samples with false labels that the model incorrectly predicts. A higher value of $R$ signifies a superior performance of the model.

$F1$ is a metric that represents the harmonic mean of precision and recall. It is commonly employed to assess a model's performance by taking into account both precision and recall. The formula for calculating $F1$ is as follows:

\begin{equation}
    F1 = 2 \times \frac{P\times R}{P+R} 
\end{equation}

\noindent where $P$ represents the precision of the model, and $R$ denotes the number of samples for which the label is false, but the model provides an incorrect prediction. A higher $F1$ score indicates superior performance of the model.

$Top-k$ is a metric used to assess the model's ability to identify vulnerable statements among the top k results it returns. The definition of $Top-k$ is as follows:

\begin{equation}
     Top-k = \frac{1}{|S_v|}\sum^{S_v}_{s=1}\delta(FRank_s \leq k)
\end{equation}

\noindent where $S_v$ represents the count of vulnerable functions, and $FRank_s$ denote the ranking assigned to the first vulnerable statement in the statement set. A higher value for $Top-k$ signifies improved performance in vulnerability localization.

$MFR$ (Mean First Ranking) is calculated as the average of the rankings assigned to the first vulnerable statement among the returned statements. The formula for calculating $MFR$ is provided below:

\begin{equation}
    MFR =  \frac{1}{|S_v|}\sum^{S_v}_{s=1}FRank_s
\end{equation}

A lower value of $MFR$ indicates superior performance in vulnerability localization.

$MAR$ (Mean Average Ranking) is calculated as the average ranking across all vulnerable statements present in the returned statements. The formula for $MAR$ is provided below:

\begin{equation}
    MAR =  \frac{1}{|S_v|}\sum^{S_v}_{s=1}\frac{1}{|N|}\sum^{N}_{i=1}Rank_{si}
\end{equation}

\noindent where $Rank_{si}$ represents the ranking of the i-th vulnerable statement within the returned statements of the s-th vulnerable function. A lower $MAR$ value indicates a superior performance in terms of vulnerability localization.

$IFA$ (Initial False Alarm) is a metric that quantifies the number of statements that are erroneously predicted as vulnerable by the models before correctly identifying the first vulnerable statement. The definition of this metric is provided below:

\begin{equation}
    IFA = \frac{1}{|S_v|}\sum^{S_v}_{s=1}(FRank_s-1)
\end{equation}

A lower value of $IFA$ indicates superior performance in vulnerability localization.

\section{Experimental Results}
\label{sec:results}

In this section, we firstly presents the experimental results and assessing the performance of \tool in terms of function-level vulnerability detection and statement-level vulnerability localization. Secondly, we evaluate the impact of Top-K statement selection on the overall performance. Thirdly, we investigate the contribution of each channel to the overall performance. Fourthly, we examine the influence of data size on both function-level vulnerability detection and statement-level vulnerability localization abilities. Fifthly, we evaluate the ability of \tool to detect different types of vulnerability. Lastly, we conduct a case study examining both successful and unsuccessful cases.

\begin{table*}[t]
\footnotesize
\setlength\tabcolsep{4pt}
\centering
\caption{
Comparison results on function-level vulnerability. The best results are highlighted in \textbf{bold} font.}
\begin{tabular}{lllllllllllllllll}
\toprule
\multirow{2}{*}{\textbf{Model}} & \multicolumn{4}{c}{\textbf{Fan \textit{et al.}}} & \multicolumn{4}{c}{\textbf{Reveal}} & \multicolumn{4}{c}{\textbf{FFMPeg+Qemu}} & \multicolumn{4}{c}{\textbf{CVEfixes}} \\
\cmidrule(lr){2-5} \cmidrule(lr){6-9} \cmidrule(lr){10-13} \cmidrule(lr){14-17}
& \textbf{Acc} & \textbf{P} & \textbf{R} & \textbf{F1} & \textbf{Acc} & \textbf{P} & \textbf{R} & \textbf{F1} & \textbf{Acc} & \textbf{P} & \textbf{R} & \textbf{F1} & \textbf{Acc} & \textbf{P} & \textbf{R} & \textbf{F1}\\
\midrule
$\rm VulDeePecker$ & 0.913 & 0.155 & 0.146 &  0.150 & 0.763 & 0.211 & 0.131 & 0.162 & 0.496 & 0.461 & 0.326 & 0.381 & 0.598 & 0.330 & 0.096 & 0.148 \\ 
$\rm SySeVR$ & 0.904 & 0.129 & 0.194 &  0.155 & 0.743 & 0.401 & 0.249 & 0.307 & 0.479 & 0.461 & 0.588 & 0.517 & 0.582 & 0.361 & 0.127 & 0.188\\ 
$\rm Devign$ & 0.957 & 0.257 & 0.143 &  0.184 & 0.875 & 0.316 & 0.367 & 0.339 & 0.569 & 0.525 & 0.647 & 0.580 & \textbf{0.615} & \textbf{0.464} & 0.076 & 0.131\\ 
$\rm Reveal$ & 0.928 & 0.270 & \textbf{0.661} &  0.383 & 0.818 & 0.316 & \textbf{0.611} & 0.416 & \textbf{0.611} & \textbf{0.555} & 0.707 & 0.622 & 0.513 & 0.412 & 0.656 & 0.506 \\
$\rm IVDetect$ & 0.696 & 0.073 & 0.600 & 0.130 & 0.808 & 0.276 & 0.556 & 0.369 & 0.573 & 0.524 & 0.576 & 0.548 & 0.601 & 0.352 & 0.085 & 0.137\\
$\rm LineVul$ & 0.972  & 0.632 & 0.436 & 0.516 & 0.847 & 0.248 & 0.519 & 0.335 & 0.541 & 0.496 & \textbf{0.909} & \textbf{0.642} & 0.496 & 0.402 & 0.672 & 0.503\\
\midrule
$\rm \tool$ & \textbf{0.977} & \textbf{0.724} & 0.522 & \textbf{0.607} & \textbf{0.922} & \textbf{0.471} & 0.394 &  \textbf{0.429} & 0.589 & 0.530 & 0.812 & 0.641 & 0.461 & 0.397 & \textbf{0.801} & \textbf{0.530} \\
$\rm \tool_{Select}$ & 0.975 & 0.678 & 0.474 & 0.558 & 0.869 & 0.294 & 0.549 &  0.383 & 0.570 & 0.520 & 0.652 & 0.578 & 0.488 &  0.391 & 0.621 & 0.480 \\
\bottomrule
\end{tabular}
\label{tab:overall_func}
\end{table*}

\begin{table*}[t]
\small
\setlength\tabcolsep{7pt}
\centering
\caption{
Comparison results on statement-level vulnerability localization. The best results are highlighted in \textbf{bold} font.}
\begin{tabular}{llllllllllll}
\toprule
\textbf{Dataset} & \textbf{Model} & \textbf{Acc} & \textbf{P} & \textbf{R} & \textbf{F1} & \textbf{MAR} & \textbf{MFR} & \textbf{IFA} & \textbf{Top-1} & \textbf{Top-3} & \textbf{Top-5}\\
\midrule
\multirow{3}{*}{Fan \textit{et al.}} & $\rm LineVul$ & N/A & N/A & N/A & N/A & 9.49 & 7.17 & 6.17 & 0.005 & 0.252 & 0.375\\
\cmidrule{2-12}
& $\rm \tool $ &  \textbf{0.983} & 0.183 & 0.338 & 0.237 & \textbf{9.08} & \textbf{6.46} & \textbf{5.46} & \textbf{0.283} & \textbf{0.484} & \textbf{0.609}\\
& $\rm \tool_{Select} $ &  0.981 & \textbf{0.185} & \textbf{0.404} & \textbf{0.253} & 9.15 & 6.63 & 5.63 & 0.243 & 0.452 & 0.572\\
\midrule
\multirow{3}{*}{CVEfixes} & $\rm LineVul$ & N/A & N/A & N/A & N/A & 9.26 & 5.62 & 4.62 & 0.221 & 0.469 & 0.585 \\
\cmidrule{2-12}
& $\rm \tool $ & 0.614 & \textbf{0.078} & \textbf{0.407} & \textbf{0.131} & 9.13 & \textbf{5.40} & \textbf{4.40} & 0.322 & \textbf{0.556} & 0.650\\
& $\rm \tool_{Select} $ & \textbf{0.647} & 0.076 & 0.352 & 0.125 & \textbf{9.08} & 5.58 & 4.58 & \textbf{0.335} & 0.551 & \textbf{0.669} \\
\bottomrule
\end{tabular}
\label{tab:overall_stat}
\end{table*}

\subsection{Comparison on function-level vulnerability detection and statement-level vulnerability localization
}
\label{sec:rq1}
Table~\ref{tab:overall_func} illustrates the comparison results of function-level vulnerability detection performance. It is worth noting that $\rm \tool_{Select}$ represents a variant of our proposed approach, and the discussion regarding it will be included in Section~\ref{sec:discussion}. In the datasets of Fan et al. and Reveal, where there is an imbalance in the proportion of positive and negative examples, the F1 metric holds more significance compared to other metrics. \revise{The results reveal that \tool outperforms other approaches and achieves state-of-the-art performance in terms of F1 on all the Fan et al., Reveal and CVEfixes datasets. Notably, \tool demonstrates a relative improvement of 17.6\%,  4.7\%, and 3.1\% in F1 on the Fan et al., CVEfixes, and Reveal datasets, respectively.} Regarding the FFMPeg+Qemu dataset, while \tool does not surpass all the baselines, its performance is closely aligned with the state-of-the-art baseline, with only a 0.2\% difference in F1. These findings demonstrate that \tool can attain performance comparable to the current state-of-the-art baselines for function-level vulnerability detection.

Table~\ref{tab:overall_stat} presents the results of statement-level vulnerability localization performance for our proposed approach and the baselines. Due to the availability of sentence-level annotation labels in the dataset by Fan et al. and CVEfixes, we display the experimental results for these two datasets. To evaluate accuracy, precision, recall, and F1, we employ the first method outlined in Section~\ref{sec:infer}, which utilizes absolute label prediction. For the remaining metrics, we utilize the second method introduced in Section~\ref{sec:infer}, employing relative scores to ensure a fair comparison between \tool and the baseline.

LineVul employs the attention score for each statement to estimate the likelihood of a statement being vulnerable, rather than directly determining its vulnerability. The metrics such as accuracy, precision, recall, and F1 are not applicable to this baseline. Instead, we present the results of our proposed \tool. A comparison between the results in Table~\ref{tab:overall_func} and Table~\ref{tab:overall_stat} reveals that the ability of \tool to detect vulnerabilities at the statement level is inferior to its ability to detect vulnerabilities at the function level. Therefore, relying solely on the statement-level predictions from \tool may not be advisable. It can be easily understood that it is quite hard to localize the vulnerable statement since there is no explicit ground-truth label for the training. Moreover, the recall of \tool is significantly higher than its precision according to the results from Table~\ref{tab:overall_stat}, indicating that \tool tends to predict statements as vulnerable at the cost of a higher false alarm rate. Comparing the performance of \tool and the baselines on the other metrics in Table~\ref{tab:overall_stat}, we observe that our proposed \tool achieves state-of-the-art performance in statement-level vulnerability localization across all metrics. Notably, there is a substantial improvement in the Top-1 metric and significant improvements in the Top-3 and Top-5 metrics. Unlike previous approaches that lack supervised signals in the attention score, our mechanism enhances vulnerability localization by providing pseudo labels during training, resulting in improved localization ability. Furthermore, the accuracy of the top 5 predictions from \tool is considerably higher than the F1 score, suggesting that our proposed \tool can effectively notify users about vulnerable statements by ranking their relative scores when the target function is predicted to be vulnerable.

In conclusion, the comprehensive experiment results shows that \tool can achieve a comparable function-level vulnerability detection performance and the state-of-the-art statement-level vulnerability localization performance compared to previous baselines, which demonstrates the effectiveness of our proposed \tool.

\begin{table*}[t]
\scriptsize
\setlength\tabcolsep{5pt}
\centering
\caption{Results of the function-level vulnerability detection performance comparison with different Top-K selection. The best results among the three variants of \tool are highlighted in \textbf{bold} font.}
\begin{tabular}{lllllllllllllllll}
\toprule
\multirow{2}{*}{\textbf{Model}} & \multicolumn{4}{c}{Fan \textit{et al.}} & \multicolumn{4}{c}{\textbf{Reveal}}  & \multicolumn{4}{c}{\textbf{FFMPeg+Qemu}}  & \multicolumn{4}{c}{\textbf{CVEfixes}} \\
\cmidrule(lr){2-5} \cmidrule(lr){6-9} \cmidrule(lr){10-13} \cmidrule(lr){14-17}
& \textbf{Acc} & \textbf{P} & \textbf{R} & \textbf{F1} & \textbf{Acc} & \textbf{P} & \textbf{R} & \textbf{F1} & \textbf{Acc} & \textbf{P} & \textbf{R} & \textbf{F1} & \textbf{Acc} & \textbf{P} & \textbf{R} & \textbf{F1}\\
\midrule
$\rm \tool_{top-1} $ & 0.976 & 0.724 & 0.481 & 0.578 & 0.913 & 0.415 & 0.423 & 0.419 & \textbf{0.589} & \textbf{0.530} & 0.812 & \textbf{0.641} & \textbf{0.515} & 0.396 & 0.527 & 0.452 \\ 
$\rm \tool_{top-3} $ & \textbf{0.977} & \textbf{0.724} & 0.522 & \textbf{0.607} & 0.894 & 0.365 & \textbf{0.471} &  0.397 & 0.575 & 0.519 & \textbf{0.824} & 0.637 & 0.461 & \textbf{0.397} & 0.801 & 0.530 \\ 
$\rm \tool_{top-5} $ & 0.974 & 0.653 & \textbf{0.524} &  0.582 & \textbf{0.922} & \textbf{0.471} & 0.394 & \textbf{0.429} & 0.576 & 0.520 & \textbf{0.824} & 0.638 & 0.446 & 0.391 & \textbf{0.827} & \textbf{0.531} \\ 
\bottomrule
\end{tabular}
\label{tab:topk_func}
\end{table*}

\begin{table*}[t]
\small
\setlength\tabcolsep{7pt}
\centering
\caption{Results of the statement-level vulnerability localization performance comparison with different Top-K selection. The best results among the three variants of \tool are highlighted in \textbf{bold} font.}
\begin{tabular}{llllllllllll}
\toprule
\textbf{Dataset} & \textbf{Model} & \textbf{Acc} & \textbf{P} & \textbf{R} & \textbf{F1} & \textbf{MAR} & \textbf{MFR} & \textbf{IFA} & \textbf{Top-1} & \textbf{Top-3} & \textbf{Top-5}\\
\midrule
\multirow{3}{*}{Fan \textit{et al.}} & $\rm \tool_{top-1} $ & \textbf{0.987} & 0.155 & 0.142 & 0.148 & 9.24 & 6.53 & 5.53 & 0.219 & 0.446 & 0.577\\
& $\rm \tool_{top-3} $ &  0.983 & \textbf{0.183} & 0.338 & \textbf{0.237} & 9.08 & 6.46 & 5.46 & 0.283 & 0.484 & \textbf{0.609}\\
& $\rm \tool_{top-5} $ & 0.979 & 0.164 & \textbf{0.380} & 0.229 & \textbf{9.02} & \textbf{6.43} & \textbf{5.43} & \textbf{0.294} & \textbf{0.497} & 0.605\\
\midrule
\multirow{3}{*}{CVEfixes} & $\rm \tool_{top-1} $ & \textbf{0.863} & \textbf{0.115} & 0.137 & 0.125 & \textbf{8.99} & \textbf{4.34} & \textbf{3.34} & \textbf{0.406} & \textbf{0.631} & \textbf{0.733}\\
& $\rm \tool_{top-3} $ & 0.614 & 0.078 & 0.407 & \textbf{0.131} & 9.13 & 5.40 & 4.40 & 0.322 & 0.556 & 0.650\\
& $\rm \tool_{top-5} $ & 0.478 & 0.073 & \textbf{0.534} & 0.128 & 9.48 & 6.02 & 5.02 & 0.303 & 0.488 & 0.621 \\
\bottomrule
\end{tabular}
\label{tab:topk_stat}
\end{table*}

\subsection{Impact of the top-k statement selection on
the performance of \tool}
\label{sec:rq2}

Table~\ref{tab:topk_func} presents the impact of the Top-K hyperparameter on the performance of function-level vulnerability detection models. It is evident that there is no universally optimal fixed value for top-k that guarantees optimal performance across all datasets. This is primarily due to the variation in the average number of vulnerable statements within functions in each dataset. Our proposed \tool employs pseudo statement-level labels as supervised signals for vulnerability localization during training. However, if the predefined top-k hyperparameter does not align with the actual number of vulnerable statements in a function, incorrect pseudo labels may be assigned. Setting a larger or smaller value of k can result in misclassifying non-vulnerable statements as vulnerable or missing some vulnerable statements, introducing noise into the model and adversely affecting performance. The results from Section~\ref{tab:stat} reveals that the average number of vulnerable statements in the Fan et al. dataset is approximately 3, which explains why the best performance is achieved when k is set to 3 in this dataset. \revise{Similarly, the model achieves the best performance when k is set to 5 on the CVEfixes dataset, where the average number of vulnerable statements exceeds 5.}

Table~\ref{tab:topk_stat} illustrates how the hyperparameter of Top-K influences the model performance of statement-level vulnerability localization. 
Before analyzing the experimental results, we must clarify that the accuracy metric in this table is meaningless due to data imbalance. All statements within non-vulnerable functions are non-vulnerable, and only a few statements within functions are vulnerable, with the rest also being non-vulnerable. Consequently, the number of non-vulnerable statements is much larger than the number of vulnerable ones, making the accuracy metric unreliable, as the model can achieve high accuracy simply by classifying all statements as non-vulnerable. We observed an interesting phenomenon: different hyperparameter values for top K lead to varying tendencies in metrics for absolute label prediction (accuracy, precision, recall, and F1) and relative scoring (MFR, MAR, IFA, Top-1, Top-3, and Top-5), as described in Section~\ref{sec:infer}. \revise{For absolute label prediction, the recall ratio increases as the k value increases. A higher k means that more statements are labeled as vulnerable during training, making the model more inclined to predict statements as vulnerable, thereby improving recall. However, a higher k negatively impacts precision, creating a trade-off when selecting an appropriate k value for absolute label prediction.}

\revise{In contrast, the impact of k on relative scoring is more complex, as the performance trend varies across different datasets. Specifically, the model performs best on the Fan et al. dataset when k is set to 5, while it achieves better results on the CVEfixes dataset when k is set to 1. This discrepancy arises because statement-level labels are derived from function-level labels, and the accuracy of these generated pseudo labels is not guaranteed. Consequently, this uncertainty affects the final performance of relative score prediction.}

In conclusion, the comprehensive experimental results demonstrate that the choice of top-k significantly impacts the performance of \tool in both function-level vulnerability detection and statement-level vulnerability localization.

\begin{table*}[t]
\scriptsize
\setlength\tabcolsep{5pt}
\centering
\caption{Results of the function-level vulnerability detection performance comparison with different channels. The best results are highlighted in \textbf{bold} font.}
\begin{tabular}{lllllllllllllllll}
\toprule
\multirow{2}{*}{\textbf{Model}} & \multicolumn{4}{c}{\textbf{Fan \textit{et al.}}} & \multicolumn{4}{c}{\textbf{Reveal}}  & \multicolumn{4}{c}{\textbf{FFMPeg+Qemu}} & \multicolumn{4}{c}{\textbf{CVEfixes}} \\
\cmidrule(lr){2-5} \cmidrule(lr){6-9} \cmidrule(lr){10-13} \cmidrule(lr){14-17}
& \textbf{Acc} & \textbf{P} & \textbf{R} & \textbf{F1} & \textbf{Acc} & \textbf{P} & \textbf{R} & \textbf{F1} & \textbf{Acc} & \textbf{P} & \textbf{R} & \textbf{F1} & \textbf{Acc} & \textbf{P} & \textbf{R} & \textbf{F1}\\
\midrule
$\rm \tool_{max} $ & 0.976 & 0.721 & 0.500 &  0.591 & \textbf{0.933} & \textbf{0.574} & 0.375 & \textbf{0.453} & 0.567 & 0.513 & \textbf{0.816} & 0.630 & \textbf{0.476} & \textbf{0.397} & 0.735 & 0.516 \\ 
$\rm \tool_{mean} $ & 0.973 & 0.660 & 0.446 & 0.532 & 0.922 & 0.470 & 0.385 & 0.423 & 0.560 & 0.509 & 0.752 & 0.608 & 0.475 & 0.395 & 0.718 & 0.509\\ 
$\rm \tool $ & \textbf{0.977} & \textbf{0.724} & \textbf{0.522} & \textbf{0.607} & 0.922 & 0.471 & \textbf{0.394} & 0.429 & \textbf{0.589} & \textbf{0.530} & 0.812 & \textbf{0.641} & 0.461 & \textbf{0.397} & \textbf{0.801} & \textbf{0.530}\\ 
\bottomrule
\end{tabular}
\label{tab:channel_func}
\end{table*}

\begin{table*}[t]
\small
\setlength\tabcolsep{7pt}
\centering
\caption{Results of the statement-level vulnerability localization performance comparison with different channels. The best results are highlighted in \textbf{bold} font.}
\begin{tabular}{lllllllllllll}
\toprule
\textbf{Dataset} & \textbf{Model} & \textbf{Acc} & \textbf{P} & \textbf{R} & \textbf{F1} & \textbf{MAR} & \textbf{MFR} & \textbf{IFA} & \textbf{Top-1} & \textbf{Top-3} & \textbf{Top-5}\\
\midrule
\multirow{3}{*}{Fan \textit{et al.}} & $\rm \tool_{max} $ & \textbf{0.984} & \textbf{0.188} & 0.299 & 0.231 & 9.17 & 6.66 & 5.66 & 0.268 & 0.481 & \textbf{0.617}\\
& $\rm \tool_{mean} $ &  0.977 & 0.155 & \textbf{0.416} & 0.226 & 9.78 & 7.27 & 6.27 & 0.224 & 0.443 & 0.586\\
& $\rm \tool $ &  0.983 & 0.183 & 0.338 & \textbf{0.237} & \textbf{9.08} & \textbf{6.46} & \textbf{5.46} & \textbf{0.283} & \textbf{0.484} & 0.609\\
\midrule
\multirow{3}{*}{CVEfixes} & $\rm \tool_{max} $ & \textbf{0.688} & 0.076 & 0.303 & 0.122 & 9.88 & 6.41 & 5.41 & 0.291 & 0.478 & 0.579\\
& $\rm \tool_{mean} $ & 0.641 & \textbf{0.078} & 0.368 & 0.128 & \textbf{9.02} & 5.44 & 4.44 & \textbf{0.326} & 0.545 & \textbf{0.682}\\
& $\rm \tool $ & 0.614 & \textbf{0.078} & \textbf{0.407} & \textbf{0.131} & 9.13 & \textbf{5.40} & \textbf{4.40} & 0.322 & \textbf{0.556} & 0.650\\
\bottomrule
\end{tabular}
\label{tab:channel_stat}
\end{table*}

\subsection{Impact of different channels on the performance of \tool
}

In this experiment, we conducted an analysis of the channels utilized in our model to determine their contribution to the performance of \tool. The results of the function-level vulnerability detection experiment using different channels are presented in Table~\ref{tab:channel_func}. It is observed that the model combining max pooling and mean pooling achieves the best performance across all datasets, except for the Reveal dataset. This outcome highlights the effectiveness of fusing max pooling and mean pooling. There are two potential explanations for why the model with only max pooling performs best in terms of the F1 metric in the dataset of Reveal. Firstly, the size of the training data could be a factor. Since the Reveal training dataset contains only around 800 vulnerable functions, the training of the model becomes unstable. As \tool has a more complex structure compared to $\rm \tool_{max}$, overfitting occurs. Secondly, the specific vulnerability type prevalent in the Reveal dataset might play a role. As explained in Section~\ref{sec:model}, the design of max pooling and mean pooling aims to capture features associated with different vulnerability types. It is possible that most vulnerabilities in the Reveal dataset align closely with the vulnerability type effectively captured by max pooling. Consequently, mean pooling may have limited contribution or even adversely affect the overall model performance. Another noteworthy finding is that $\rm \tool_{max}$ performs very similarly to \tool, while $\rm \tool_{mean}$ exhibits considerably worse performance than $\rm \tool_{max}$. This discrepancy may be attributed to the fact that many vulnerabilities are related to specific keywords within statements, and these features are effectively captured by the max pooling mechanism. \revise{Listing~\ref{lst:given_max_example} presents an illustrative example. \tool effectively detects and localizes use-after-free vulnerabilities by identifying the repeated occurrence of keywords like `buffer,' which has already been deleted in a previous statement. In contrast, mean pooling averages the high-dimensional representation vectors for each token in the statement, potentially diluting these keyword features and reducing the model's performance.}

\lstset{language=C}
\begin{lstlisting}[caption={An example for vulnerabilty related to specific keywords within statements.},label={lst:given_max_example}]
#include <iostream>
#include <fstream>

void processFile(const std::string& filename) {
    char* buffer = new char[1024]; 
    std::ifstream file(filename, std::ios::binary);
    
    if (!file) {
        std::cerr << "Failed to open the file." << std::endl;
        delete[] buffer;
        return;
    }

    file.read(buffer, 1024);
    std::cout << "File content read successfully." << std::endl;

    delete[] buffer;
    std::cout << "Buffer memory freed." << std::endl;

    std::cout << "First byte of file content (post-delete): " << buffer[0] << std::endl;
}
\end{lstlisting}

Table~\ref{tab:channel_stat} presents the experimental results for statement-level vulnerability localization using different channels. The results demonstrate that our \tool performs the best across most metrics, showcasing the effectiveness of the fusion of max pooling and mean pooling. \revise{Unlike the conclusion drawn from Table~\ref{tab:channel_func}, the performance of max channel and mean channel in statement-level vulnerability localization varies across different datasets. This discrepancy may be attributed to the differences in CVE types among the datasets.}

\wcgu{In conclusion, the comprehensive experimental results validate the effectiveness of both max pooling and mean pooling in the proposed \tool. This combination significantly enhances the ability of function-level vulnerability detection and statement-level vulnerability localization.}

\begin{table*}[t]
\small
\setlength\tabcolsep{3.5pt}
\centering
\caption{
Comparison results on function-level vulnerability detection and statement-level vulnerability localization with different sizes of training data in the dataset of Fan et al.. The best results are highlighted in \textbf{bold} font.}
\begin{tabular}{llllllllllllllll}
\toprule
\multirow{2}{*}{\textbf{Dataset}} & \multirow{2}{*}{\textbf{Model}} & \multicolumn{4}{c}{\textbf{Function-level Detection}} & \multicolumn{10}{c}{\textbf{Statement-level Localization}} \\
\cmidrule(lr){3-6} \cmidrule(lr){7-16}
& & \textbf{Acc} & \textbf{P} & \textbf{R} & \textbf{F1} & \textbf{Acc} & \textbf{P} & \textbf{R} & \textbf{F1} & \textbf{MAR} & \textbf{MFR} & \textbf{IFA} & \textbf{Top-1} & \textbf{Top-3} & \textbf{Top-5}\\
\midrule

\multirow{4}{*}{Fan et. al} & $\rm \tool_{10\%}$ & 0.969 & 0.580 & 0.265 & 0.364 & 0.991 & 0.074 & 0.014 & 0.024 & 9.33 & 6.73 & 5.73 & 0.235 & 0.444 & 0.585\\
& $\rm \tool_{20\%}$ & 0.971  & 0.610 & 0.398 & 0.482 & 0.987 & 0.137 & 0.118 & 0.127 & \textbf{9.07} & 6.48 & 5.48 & 0.227 & 0.431 & 0.583\\
& $\rm \tool_{50\%}$ & 0.971 & 0.591 & 0.521 & 0.554 & 0.985 & 0.170 & 0.217 & 0.191 & 9.33 & 6.64 & 5.64 & 0.260 & 0.460 & 0.593 \\
& $\rm \tool$ & \textbf{0.977}  & \textbf{0.724} & \textbf{0.522} & \textbf{0.607} & \textbf{0.983} & \textbf{0.183} & \textbf{0.338} & \textbf{0.237} & 9.08 & \textbf{6.46} & \textbf{5.46} & \textbf{0.283} & \textbf{0.484} & \textbf{0.609}\\
\midrule

\multirow{4}{*}{CVEfixes} & $\rm \tool_{10\%}$ & \textbf{0.596} & \textbf{0.422} & 0.170 & 0.242 & \textbf{0.920} & 0.075 & 0.011 & 0.018 & 9.80 & 8.51 & 7.51 & 0.179 & 0.358 & 0.471 \\
& $\rm \tool_{20\%}$ & 0.534 & 0.380 & 0.357 & 0.368 & 0.872 & 0.079 & 0.073 & 0.076 & 9.72 & 7.66 & 6.66 & 0.191 & 0.389 & 0.520 \\
& $\rm \tool_{50\%}$ & 0.543 & 0.400 & 0.403 & 0.401 & 0.851 & \textbf{0.102} & 0.139 & 0.118 & \textbf{9.13} & 5.43 & 4.43 & \textbf{0.328} & \textbf{0.568} & \textbf{0.661} \\
& $\rm \tool$ & 0.461 & 0.397 & \textbf{0.801} & \textbf{0.530} & 0.614 & 0.078 & \textbf{0.407} & \textbf{0.131} & \textbf{9.13} & \textbf{5.40} & \textbf{4.40} & 0.322 & 0.556 & 0.650\\
\bottomrule
\end{tabular}
\label{tab:ratio}
\end{table*}

\subsection{The influence of the training data size to the performance of \tool}

In this experiment, we assess the impact of training data size on the performance of our proposed \tool for function-level vulnerability detection and statement-level vulnerability localization. Since only the dataset of Fan et al. includes statement-level labels, we exclusively evaluate our \tool using this dataset. To thoroughly examine the influence of training data size on model performance, we randomly select 10\%, 20\%, 50\%, and 100\% of the data from the training dataset to construct new training datasets. Table~\ref{tab:ratio} presents the experimental results for function-level vulnerability detection and statement-level vulnerability localization across different training data sizes. As expected, the model's performance in function-level vulnerability detection consistently improves as the training data size increases. Interestingly, even though the data lacks statement-level labels, the model's performance in statement-level vulnerability localization also improves as the training data size increases. These findings demonstrate that our proposed \tool becomes more proficient at locating vulnerability statements as the training data size grows, even without explicit annotations. However, the performance improvement of \tool on absolute label prediction and relative scores, as introduced in Section~\ref{sec:infer}, varies with the increase in training data size. Specifically, the performance improvement in absolute label prediction, measured by accuracy, precision, recall, and F1, is substantial with larger training data sizes. Conversely, \tool maintains most of its performance in relative scores metrics even with limited training data, and the performance improvement in relative scores is not as rapid as the improvement in absolute label prediction with increasing training data size. \revise{In the CVEfixes dataset, the performance of relative score predictions using 50\% of the data is nearly identical to that achieved with 100\% of the data.}

In summary, increasing the size of the training data can enhance the performance of function-level vulnerability detection and statement-level vulnerability localization, even without any additional information about the vulnerable statements in the data.

\begin{table*}[t]
\setlength\tabcolsep{8pt}
\centering
\caption{Detection results for different CWE vulnerabilities with our proposed \tool.}
\begin{tabular}{lllll}
\toprule
\textbf{CWE-ID} & \textbf{Description} & \textbf{Rank} & \textbf{TPR} & \textbf{Proportion}\\
\midrule
CWE-787 & Out-of-bounds Write & 1 & 50.0\% & 7/14\\
CWE-416 & Use After Free & 4 & 61.5\% & 8/13 \\
CWE-20 & Improper Input Validation & 6 & 56.6\% & 61/108\\
CWE-125 & Out-of-bounds Read & 7 & 54.2\% & 13/24 \\
CWE-476 & NULL Pointer Dereference & 12 & 55.6\% & 5/9 \\

CWE-190 & Integer Overflow or Wraparound & 14 & 77.8\% & 14/18 \\
CWE-119 & Improper Restriction of Operations within the Bounds of & 17 & 56.0\% & 70/125 \\
& a Memory Buffer & & \\
CWE-362 & Concurrent Execution using Shared Resource with Improper & 21 & 52.4\% & 11/21 \\
& Synchronization ('Race Condition') & & \\
\hdashline
CWE-284 & Improper Access Control & N/A & 37.5\% & 3/8 \\
CWE-189 & Numeric Errors & N/A & 52.6\% & 10/19 \\
CWE-732 & Incorrect Permission Assignment for Critical Resource & N/A & 57.1\% & 4/7 \\
CWE-254 & 7PK - Security Features & N/A & 44.4\% & 4/9 \\
CWE-200 & Exposure of Sensitive Information to an Unauthorized Actor & N/A & 42.9\% & 12/28 \\
CWE-415 &  Double Free & N/A & 71.4\% & 5/7 \\
CWE-399 & Resource Management Errors & N/A & 48.7\% & 19/39 \\
\midrule
Total & & &  54.8\% & 246/449 \\ 
\bottomrule
\end{tabular}
\label{tab:type}
\end{table*}

\subsection{The detection ability of \tool for different types of CWE vulnerabilities}

Table~\ref{tab:type} presents the detection results of our proposed \tool for \yl{different types of vulnerabilities from CWE}. To ensure meaningful data analysis, we filtered out vulnerability types with a small number of occurrences and retained only those with at least five instances from the same test dataset, which was also used in previous experiments. The CWE has released the 2023 Top 25 Most Dangerous Software Weaknesses on their website~\cite{CWE}. The rank metric in the table indicates the CWE's ranking in their list. In cases where the vulnerability types in our test dataset are not included in the Top 25 list, we denote them as ``N/A'' in the rank metric. Additionally, the TPR metric in the table represents the true positive rate, indicating the percentage of \yl{vulnerabilities successfully detected by the model}.

\yl{The results reveal an interesting finding: the detection ability of \tool does not vary significantly between different types of CWE vulnerability, with a TPR of approximately 56\% for most CWE types. }However, certain vulnerabilities, such as CWE-415 and CWE-284, exhibit notably high or low TPRs. It is important to note that the limited number of vulnerabilities makes it difficult to conclusively determine whether \tool performs well or poorly in these specific types. Nevertheless, some exceptions stand out. Notably, \tool demonstrates a strong detection ability for vulnerabilities classified as CWE-190, successfully identifying almost all instances of this type. Conversely, \tool displays a poor detection ability for vulnerabilities categorized as CWE-200 and CWE-399, as it only identifies less than half of the vulnerabilities within these types. We believe that the accuracy of detecting different types of vulnerabilities may be influenced by the difficulty of identifying their patterns. Our findings indicate that the detection accuracy for use-after-free, integer overflow or wraparound, and double-free vulnerabilities is higher than for other types. These vulnerabilities have relatively fixed patterns, making them easier to detect by conventional static analysis methods. And our proposed approach is likely to detect these patterns more effectively as well.

In conclusion, \tool demonstrates comparable detection abilities for \yl{most types of vulnerabilities}. Nevertheless, it exhibits superior performance in detecting vulnerabilities related to Integer Overflow or Wraparound, while its effectiveness is relatively weaker in identifying vulnerabilities associated with Exposure of Sensitive Information to an Unauthorized Actor and Resource Management Errors.

\subsection{Case Study}

Listing~\ref{lst:succ_case1} shows a successfully detected vulnerability associated with CWE-20, which is related to Improper Input Validation. In this code snippet, the \texttt{memcpy} function is invoked on lines 6 and 8. However, the input length for this function is not validated. If the length of the data pointed to by the \texttt{smac} or \texttt{alt\_smac} pointer exceeds the size of \texttt{av\_smac\_size} or \texttt{alt\_av\_smac\_size}, the \texttt{memcpy} function will copy data beyond the boundary of the target array, resulting in a buffer overflow. In this case, the lack of comprehensive verification of the input data's validity may lead to unpredictable system behavior or a system crash, especially when processing network packets. This case indicates that \tool can capture the feature of missing input validation before the execution of the \texttt{memcpy} function during contextual learning and issue warnings about these two statements.

\lstset{language=C}
\begin{lstlisting}[caption={Successfully detected vulnerability associated with CWE-20},label={lst:succ_case1}]
int ib_update_cm_av(struct ib_cm_id *id, const u8 *smac, const u8 *alt_smac)
{
        struct cm_id_private *cm_id_priv;
        cm_id_priv = container_of(id, struct cm_id_private, id);
        if (smac != NULL)
                memcpy(cm_id_priv->av.smac, smac, sizeof(cm_id_priv->av.smac));
        if (alt_smac != NULL)
                memcpy(cm_id_priv->alt_av.smac, alt_smac,
                       sizeof(cm_id_priv->alt_av.smac));
        return 0;
}
\end{lstlisting}

Listing~\ref{lst:succ_case2} shows another successfully detected vulnerability associated with CWE-200, which relates to the exposure of sensitive information to unauthorized actors. In this code snippet, line 26 logs error information when \texttt{valid\_path} does not exist. However, the user email address is included in the \texttt{valid\_path} variable on lines 17 and 18, so user information will also be logged. This recording can expose sensitive user information, especially if the logs are accessible to unauthorized personnel. In this case, \tool successfully captures sensitive tokens like the user email and detects the sensitive information logging behavior, issuing a warning about this vulnerability.

\lstset{language=C}
\begin{lstlisting}[caption={Successfully detected vulnerability associated with CWE-200},label={lst:succ_case2}]
void WallpaperManagerBase::GetCustomWallpaperInternal(
    const AccountId& account_id,
    const WallpaperInfo& info,
    const base::FilePath& wallpaper_path,
    bool update_wallpaper,
    const scoped_refptr<base::SingleThreadTaskRunner>& reply_task_runner,
    MovableOnDestroyCallbackHolder on_finish,
    base::WeakPtr<WallpaperManagerBase> weak_ptr) {
  base::FilePath valid_path = wallpaper_path;
  if (!base::PathExists(wallpaper_path)) {
    valid_path = GetCustomWallpaperDir(kOriginalWallpaperSubDir);
    valid_path = valid_path.Append(info.location);
  }

  if (!base::PathExists(valid_path)) {
    LOG(ERROR) << "Failed to load custom wallpaper from its original fallback "
                  "file path: " << valid_path.value();
     const std::string& old_path = account_id.GetUserEmail();  // Migrated
     valid_path = GetCustomWallpaperPath(kOriginalWallpaperSubDir,
                                         WallpaperFilesId::FromString(old_path),
                                         info.location);
   }

   if (!base::PathExists(valid_path)) {
    LOG(ERROR) << "Failed to load previously selected custom wallpaper. "
               << "Fallback to default wallpaper. Expected wallpaper path: "
               << wallpaper_path.value();
     reply_task_runner->PostTask(
         FROM_HERE,
        base::Bind(&WallpaperManagerBase::DoSetDefaultWallpaper, weak_ptr,
                   account_id, base::Passed(std::move(on_finish))));
   } else {
     reply_task_runner->PostTask(
         FROM_HERE, base::Bind(&WallpaperManagerBase::StartLoad, weak_ptr,
                              account_id, info, update_wallpaper, valid_path,
                              base::Passed(std::move(on_finish))));
  }
}
\end{lstlisting}

\subsection{Error Analysis}

Despite \tool's improved performance over previous vulnerability localization methods, certain cases remain undetected or mislocalized. We have selected two such cases for error analysis.

\subsubsection{Missing Implementation of Invoked Function}
Listing~\ref{lst:failed_case1} is a failed detection example associated with CWE-787. This code snippet contains a potential Out-of-bounds Write issue, which allowed a remote attacker who had compromised the renderer process to perform an out-of-bounds memory write via a crafted HTML page. However, this vulnerability is related to the implementation of the invoked function \texttt{CopyMetafileDataToSharedMem}, and such implementation details are not included in the code snippet. The absence of this key information caused \tool to fail in detecting this vulnerability.

\lstset{language=C}
\begin{lstlisting}[caption={
Case of failed vulnerability detection associated with CWE-787},label={lst:failed_case1}]
bool PrintRenderFrameHelper::PrintPagesNative(blink::WebLocalFrame* frame,
                                              int page_count) {
  const PrintMsg_PrintPages_Params& params = *print_pages_params_;
  const PrintMsg_Print_Params& print_params = params.params;

  std::vector<int> printed_pages = GetPrintedPages(params, page_count);
  if (printed_pages.empty())
    return false;

  PdfMetafileSkia metafile(print_params.printed_doc_type);
  CHECK(metafile.Init());

  PrintHostMsg_DidPrintDocument_Params page_params;
  PrintPageInternal(print_params, printed_pages[0], page_count, frame,
                    &metafile, &page_params.page_size,
                    &page_params.content_area);
  for (size_t i = 1; i < printed_pages.size(); ++i) {
    PrintPageInternal(print_params, printed_pages[i], page_count, frame,
                      &metafile, nullptr, nullptr);
  }

  FinishFramePrinting();

   metafile.FinishDocument();

  if (!CopyMetafileDataToSharedMem(metafile,
                                   &page_params.metafile_data_handle)) {
     return false;
   }

  page_params.data_size = metafile.GetDataSize();
  page_params.document_cookie = print_params.document_cookie;
#if defined(OS_WIN)
  page_params.physical_offsets = printer_printable_area_.origin();
#endif
  Send(new PrintHostMsg_DidPrintDocument(routing_id(), page_params));
  return true;
}
\end{lstlisting}

\subsubsection{Missing External Knowledge}
Listing~\ref{lst:failed_case2} illustrates another type of detection failure associated with CWE-241. CWE-241 pertains to the improper handling of unexpected data types, where a product does not correctly handle an element that is not of the expected type. In this code snippet, \texttt{raw\_headers.push\_back('\textbackslash 0')} is used to terminate each line, with \texttt{\textbackslash 0\textbackslash 0} appended as the terminator at the end. This approach does not comply with the HTTP protocol specification and can cause the code parsing HTTP header information to misinterpret and misprocess the headers. Sometimes, HTTP header information may include user input data. If this data contains the \texttt{\textbackslash 0} character, using \texttt{\textbackslash 0} as the terminator might lead to injection attacks, as the parser may prematurely terminate parsing, resulting in incomplete or incorrect parsing outcomes. However, without the specific knowledge about the HTTP protocol, \tool failed to recognize this as a vulnerability, leading to the detection failure.

\lstset{language=C}
\begin{lstlisting}[caption={
Case of failed vulnerability detection associated with CWE-241},label={lst:failed_case2}]
std::string HttpUtil::AssembleRawHeaders(const char* input_begin,
                                         int input_len) {
  std::string raw_headers;
  raw_headers.reserve(input_len);

  const char* input_end = input_begin + input_len;

  int status_begin_offset = LocateStartOfStatusLine(input_begin, input_len);
  if (status_begin_offset != -1)
    input_begin += status_begin_offset;

  const char* status_line_end = FindStatusLineEnd(input_begin, input_end);
  raw_headers.append(input_begin, status_line_end);


  CStringTokenizer lines(status_line_end, input_end, "\r\n");

  bool prev_line_continuable = false;

  while (lines.GetNext()) {
    const char* line_begin = lines.token_begin();
    const char* line_end = lines.token_end();

    if (prev_line_continuable && IsLWS(*line_begin)) {
      raw_headers.push_back(' ');
       raw_headers.append(FindFirstNonLWS(line_begin, line_end), line_end);
     } else {
      raw_headers.push_back('\0');

       raw_headers.append(line_begin, line_end);

      prev_line_continuable = IsLineSegmentContinuable(line_begin, line_end);
     }
   }

  raw_headers.append("\0\0", 2);
   return raw_headers;
 }

\end{lstlisting}
\section{Discussion}
\label{sec:discussion}

In this section, we first explore the impact of different channel fusion strategies on model performance. Then, we examine software developers' attitudes toward vulnerability detection and localization tools.

\subsection{Performance of using different channel fusion strategies}

In this subsection, we explore the impact of employing various channel fusion strategies on model performance. In our proposed method, we combine the scores from two linear classifiers using a weighted sum to create a single prediction score. We compare this approach with another strategy, denoted as $\rm \tool_{Select}$, where we select the maximum score from the two linear classifiers as the prediction score. Analyzing the results presented in Table~\ref{tab:overall_func}, we observe that the strategy that fuses prediction scores from both max pooling and mean pooling channels significantly outperforms the strategy of selecting the maximum score from the two channels. This finding suggests that focusing solely on either max pooling or mean pooling features is insufficient for effective vulnerability detection; it is essential to consider prediction results from both channels simultaneously.

\revise{The results of statement-level vulnerability localization, presented in Table~\ref{tab:overall_stat}, exhibit a different trend compared to those in Table~\ref{tab:channel_stat}. The effectiveness of different strategies varies across datasets. Specifically, on the dataset from Fan et al., $\rm \tool_{Select}$ outperforms $\rm \tool$ in absolute label prediction (Accuracy, Precision, Recall, and F1) but underperforms in relative score ranking (MAR, MFR, IFA, Top-1, Top-3, Top-5). In contrast, on the CVEfixes dataset, $\rm \tool_{Select}$ performs worse in absolute label prediction but achieves comparable performance to $\rm \tool$ in relative score ranking.}

\subsection{Human assessment of the effectiveness of vulnerability localization}

\begin{figure}[t]
\centering
\subfloat[\scriptsize Results from the questionnaire on confidence in recognizing vulnerability]{
\includegraphics [width=4.8cm]{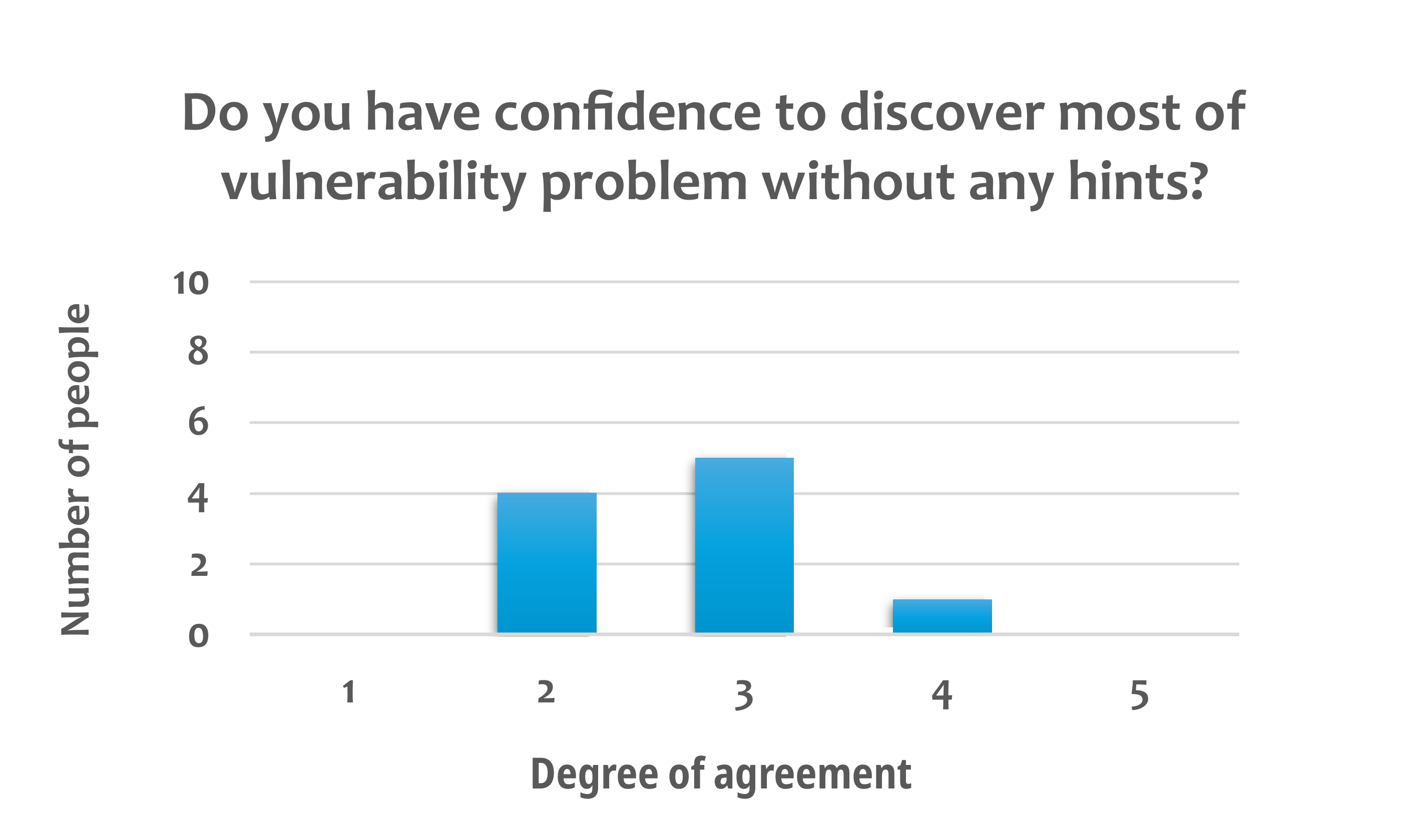}}
\hspace{0.1cm}
\subfloat[\scriptsize Results from the questionnaire regarding users' confidence in vulnerability localization]{
\includegraphics [width=4.8cm]{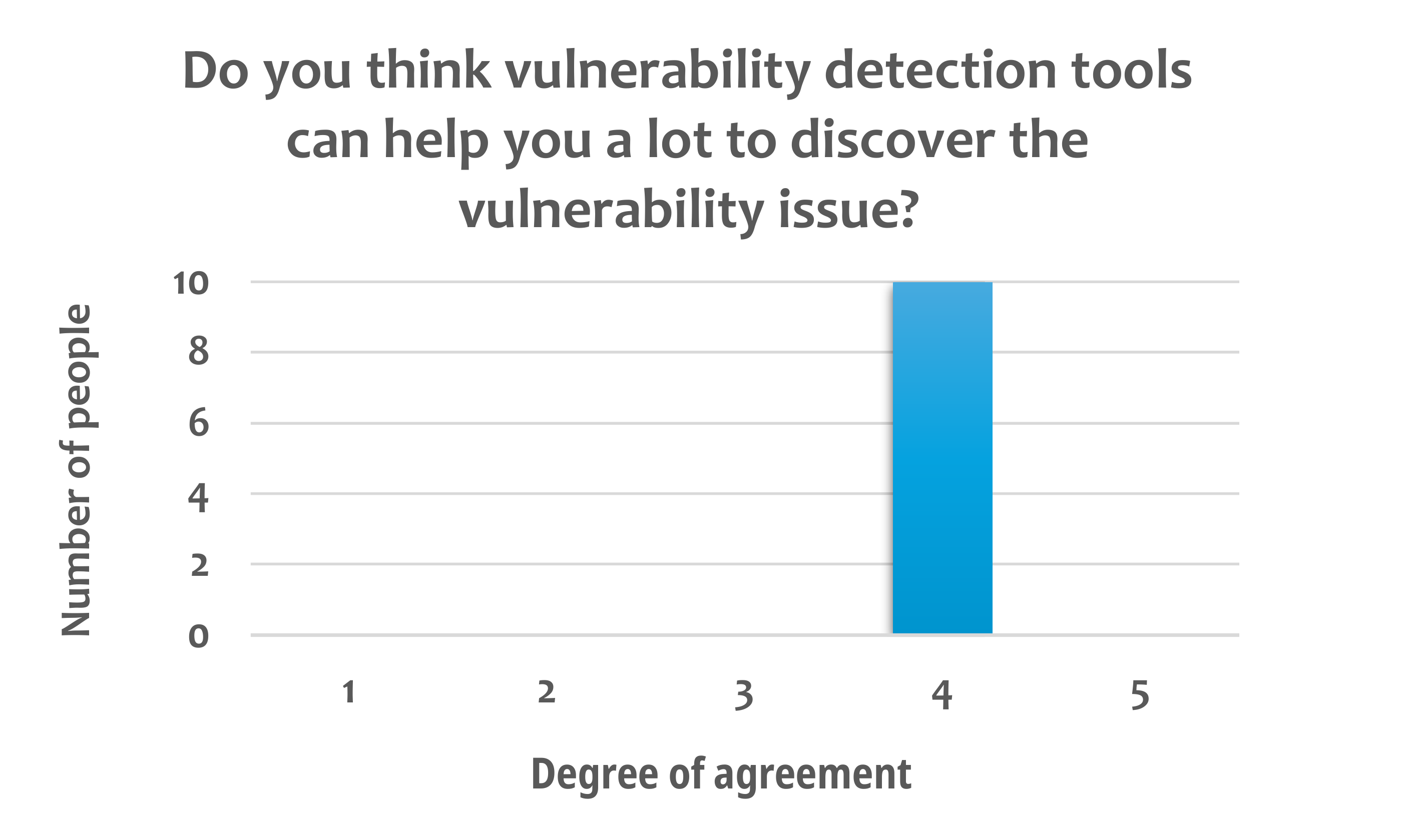}}
\hspace{0.1cm}
\subfloat[\scriptsize Results from the questionnaire assessing the effectiveness of vulnerability detection tools]{
\includegraphics [width=4.8cm]{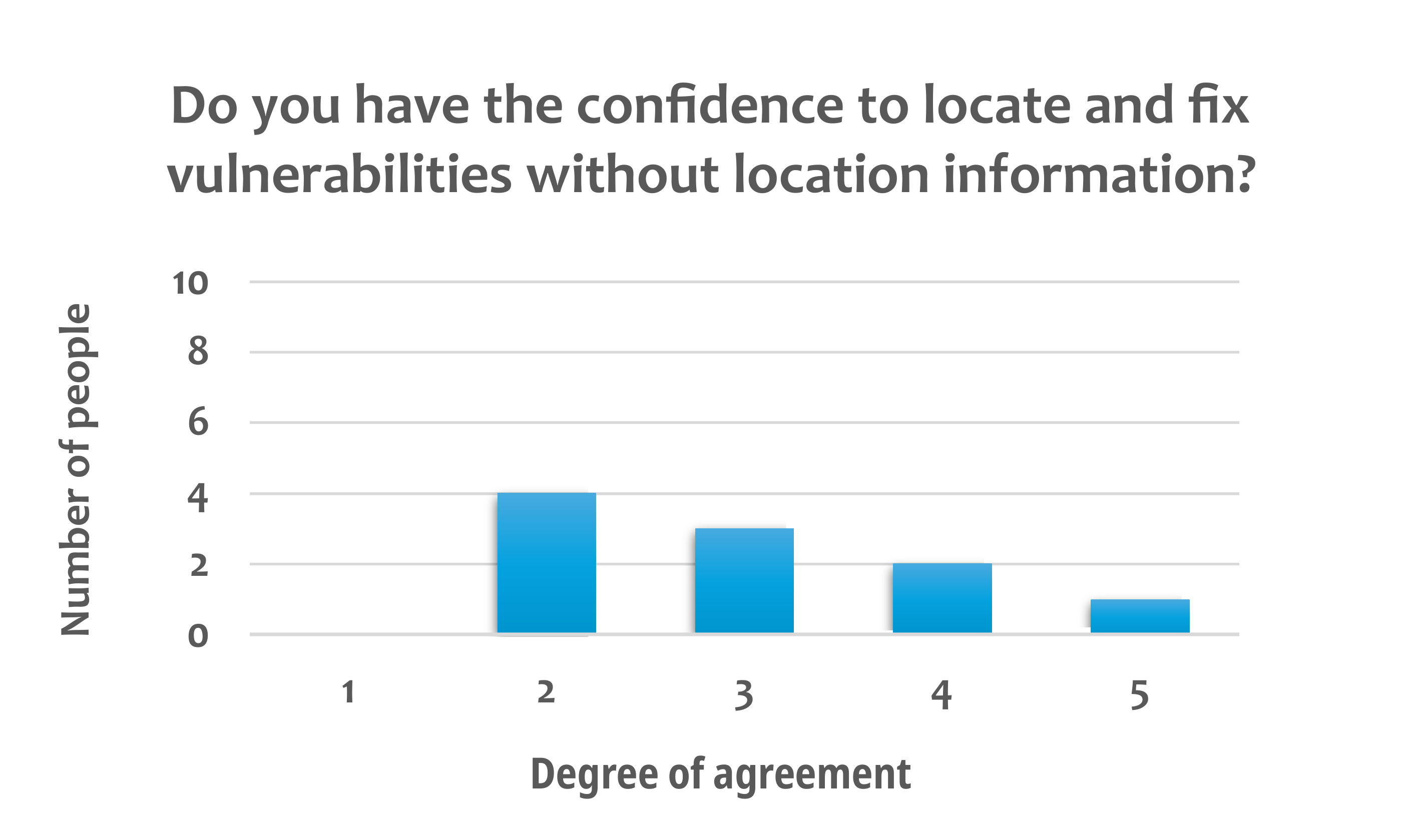}}
\quad
\subfloat[\scriptsize Results from the questionnaire on the necessity of vulnerability localization]{
\includegraphics [width=4.8cm]{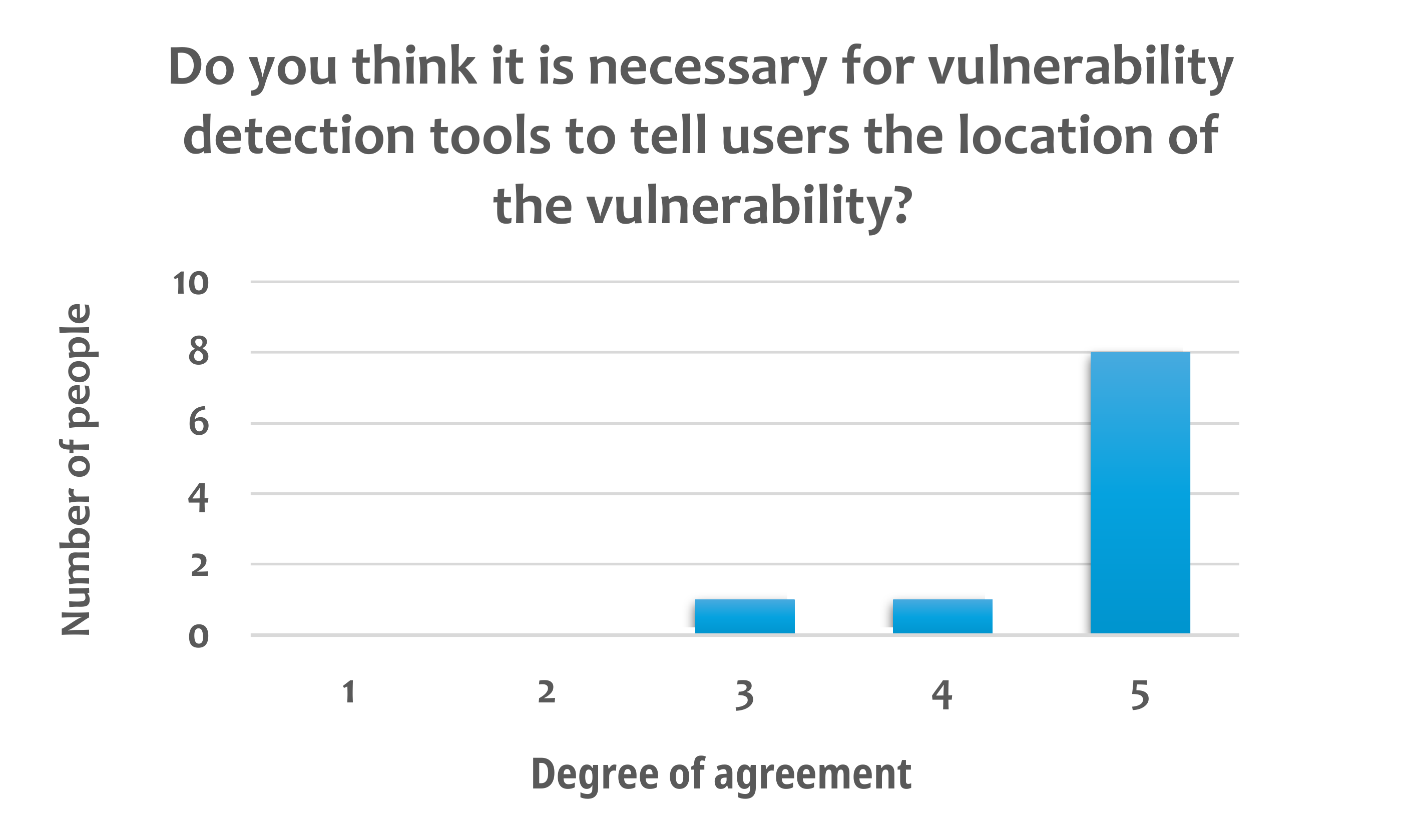}}
\hspace{0.1cm}
\subfloat[\scriptsize Results from the questionnaire regarding the effectiveness of vulnerability localization]{
\includegraphics [width=4.8cm]{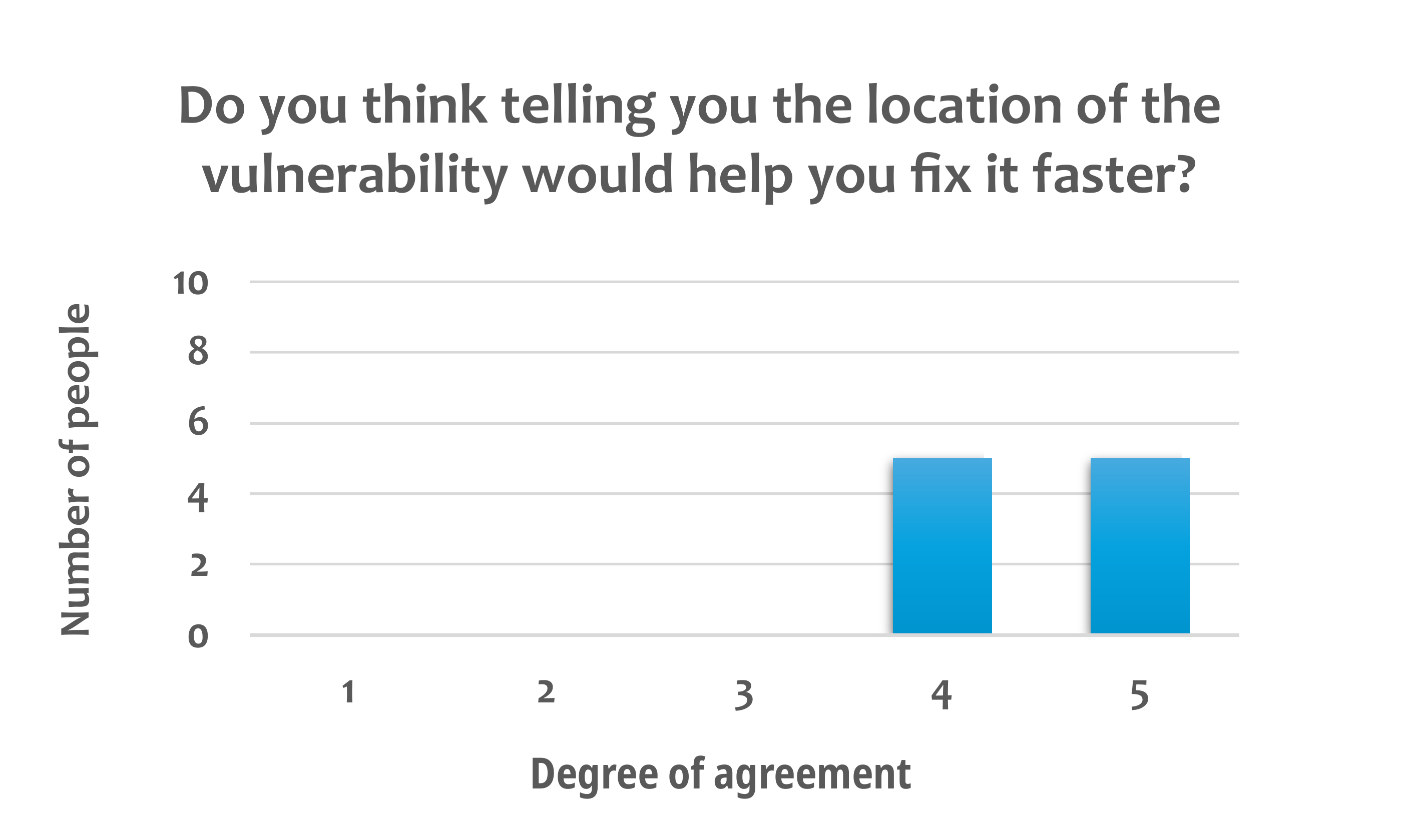}}
\hspace{0.1cm}
\subfloat[\scriptsize Results from the questionnaire evaluating the effectiveness of vulnerability localization in Case 1]{
\includegraphics [width=4.8cm]{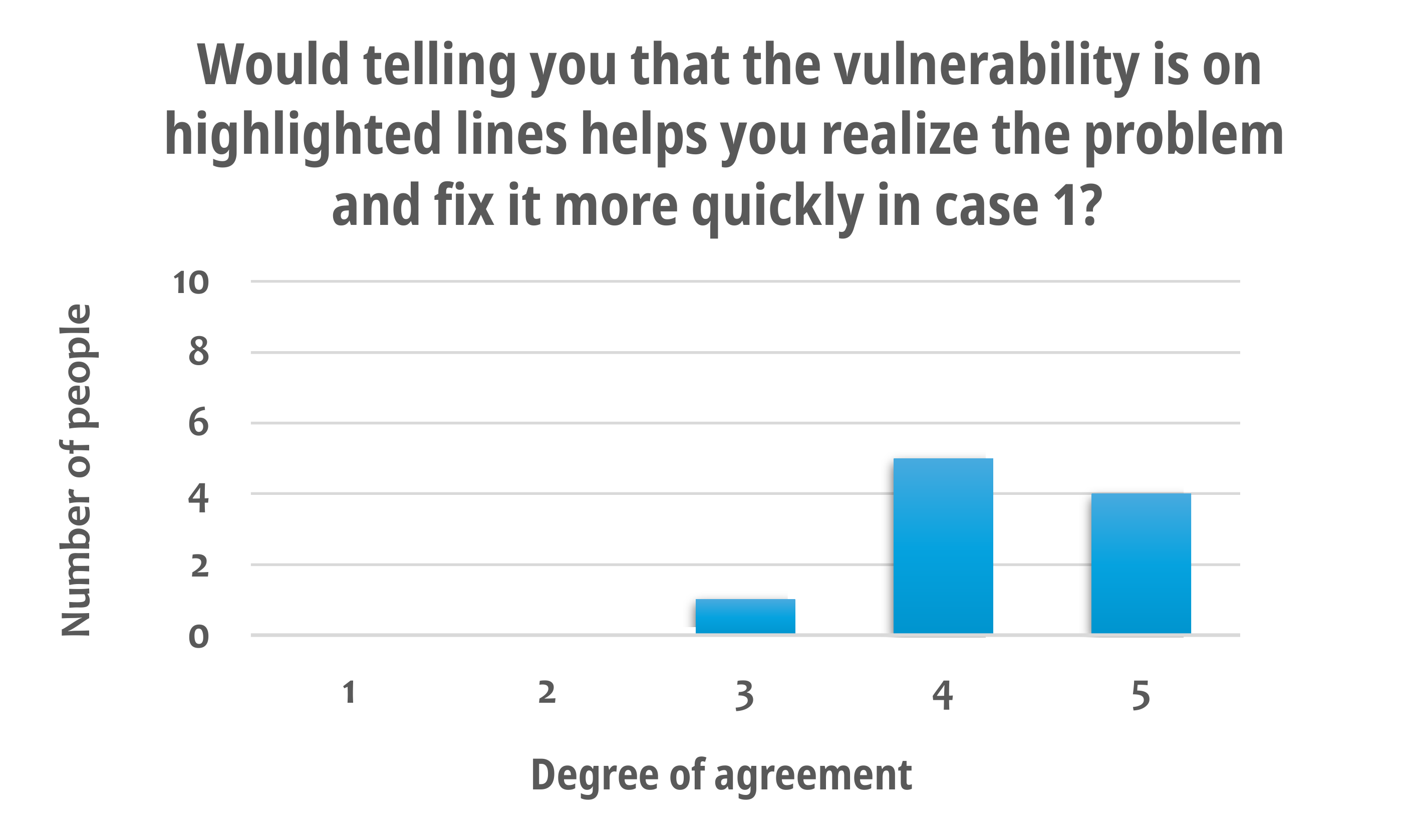}}
\quad
\subfloat[\scriptsize Results from the questionnaire evaluating the effectiveness of vulnerability localization in Case 2]{
\includegraphics [width=4.8cm]{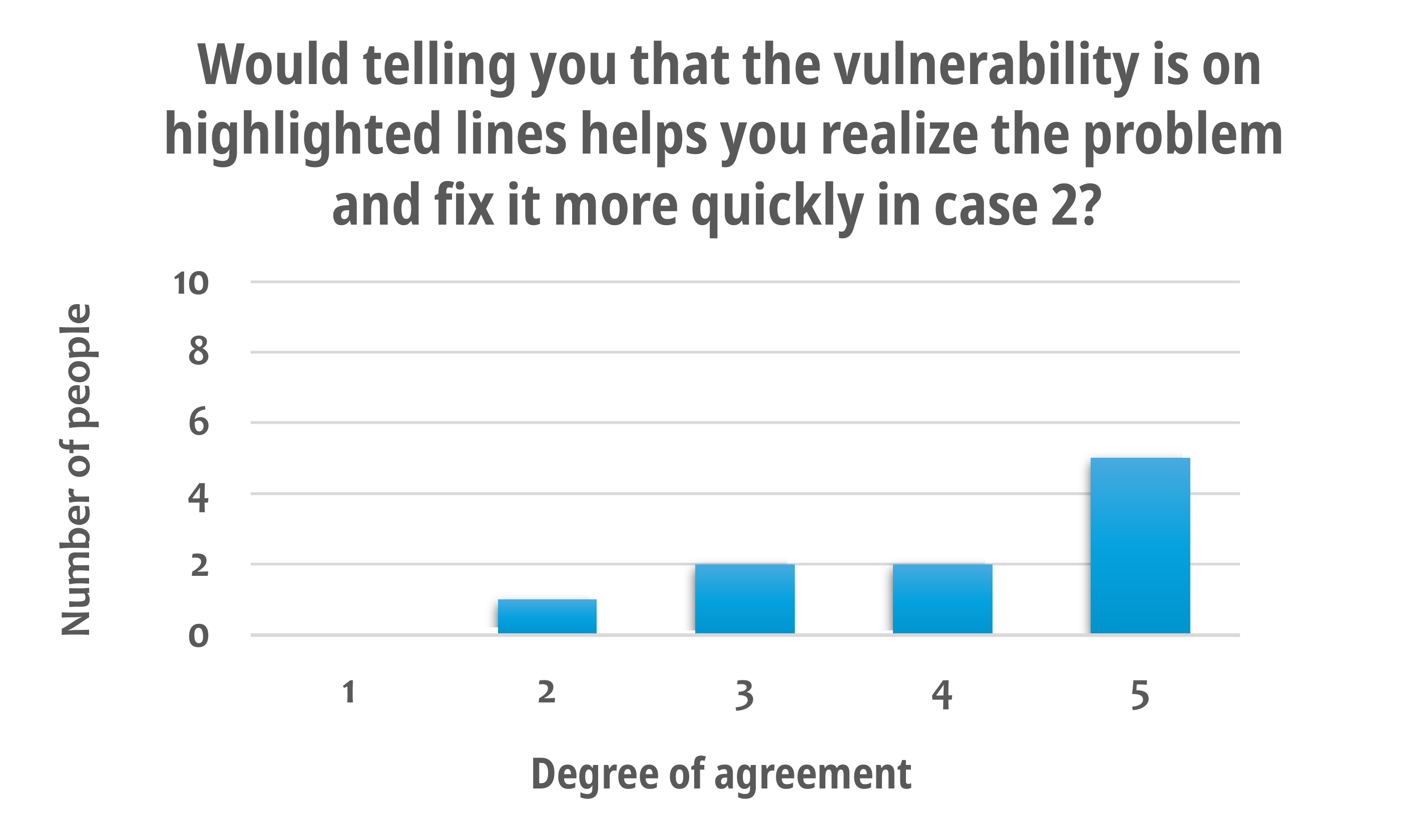}}
\hspace{0.1cm}
\subfloat[\scriptsize Results from the questionnaire evaluating the effectiveness of vulnerability localization in Case 3]{
\includegraphics [width=4.8cm]{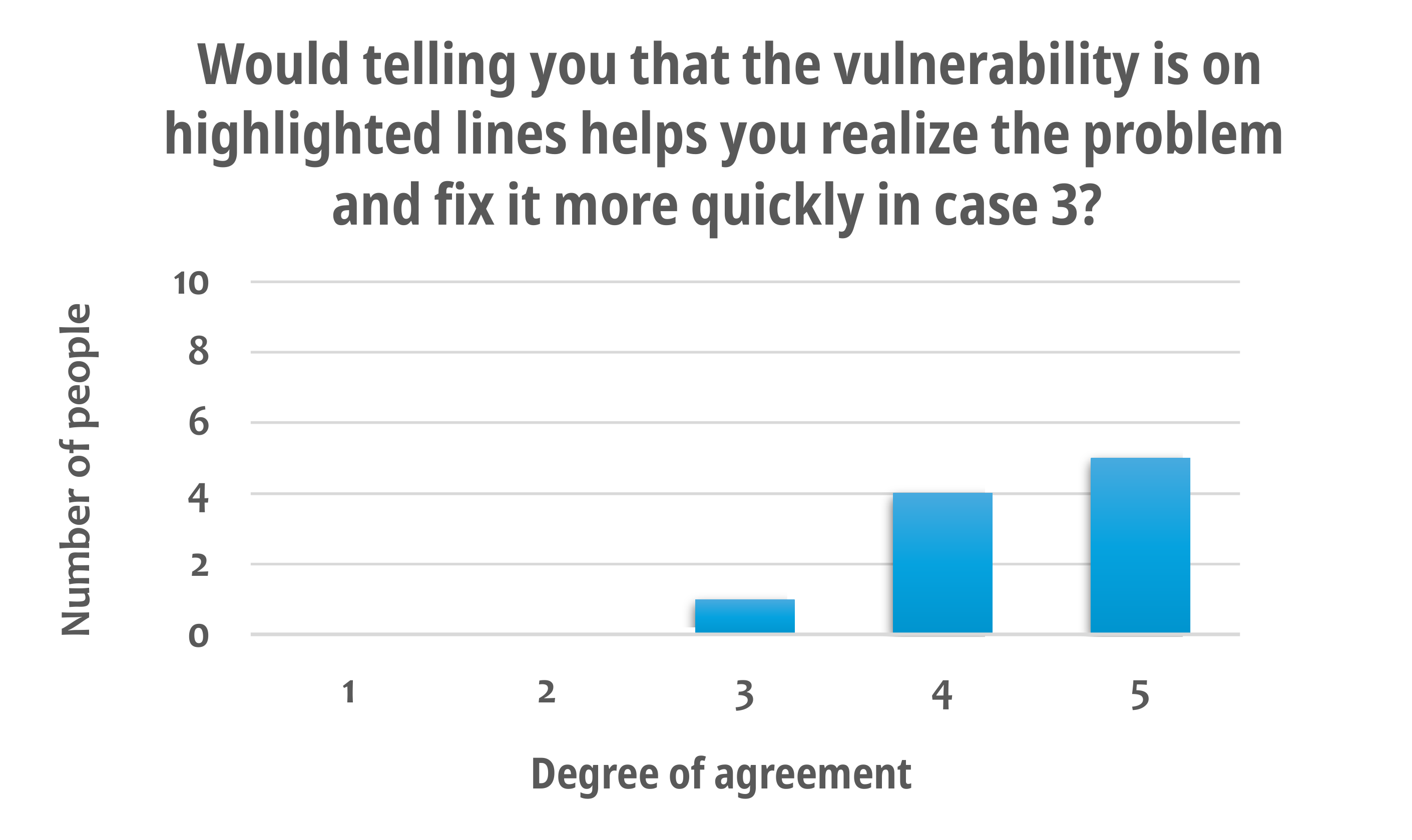}}
\hspace{0.1cm}
\subfloat[\scriptsize Results from the questionnaire evaluating the effectiveness of vulnerability localization in Case 4]{
\includegraphics [width=4.8cm]{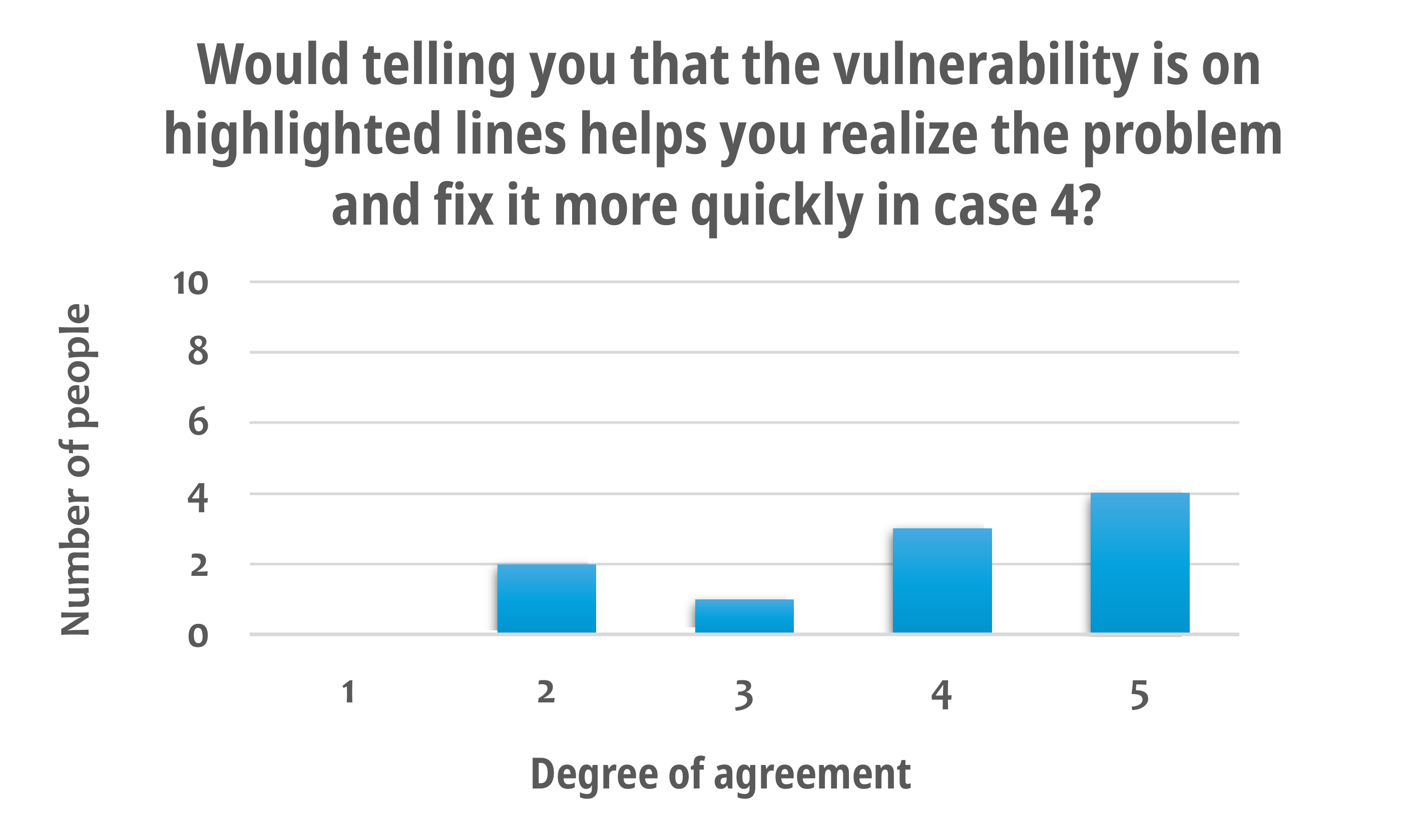}}
\hspace{0.1cm}
\subfloat[\scriptsize Results from the questionnaire evaluating the effectiveness of vulnerability localization in Case 5]{
\includegraphics [width=4.8cm]{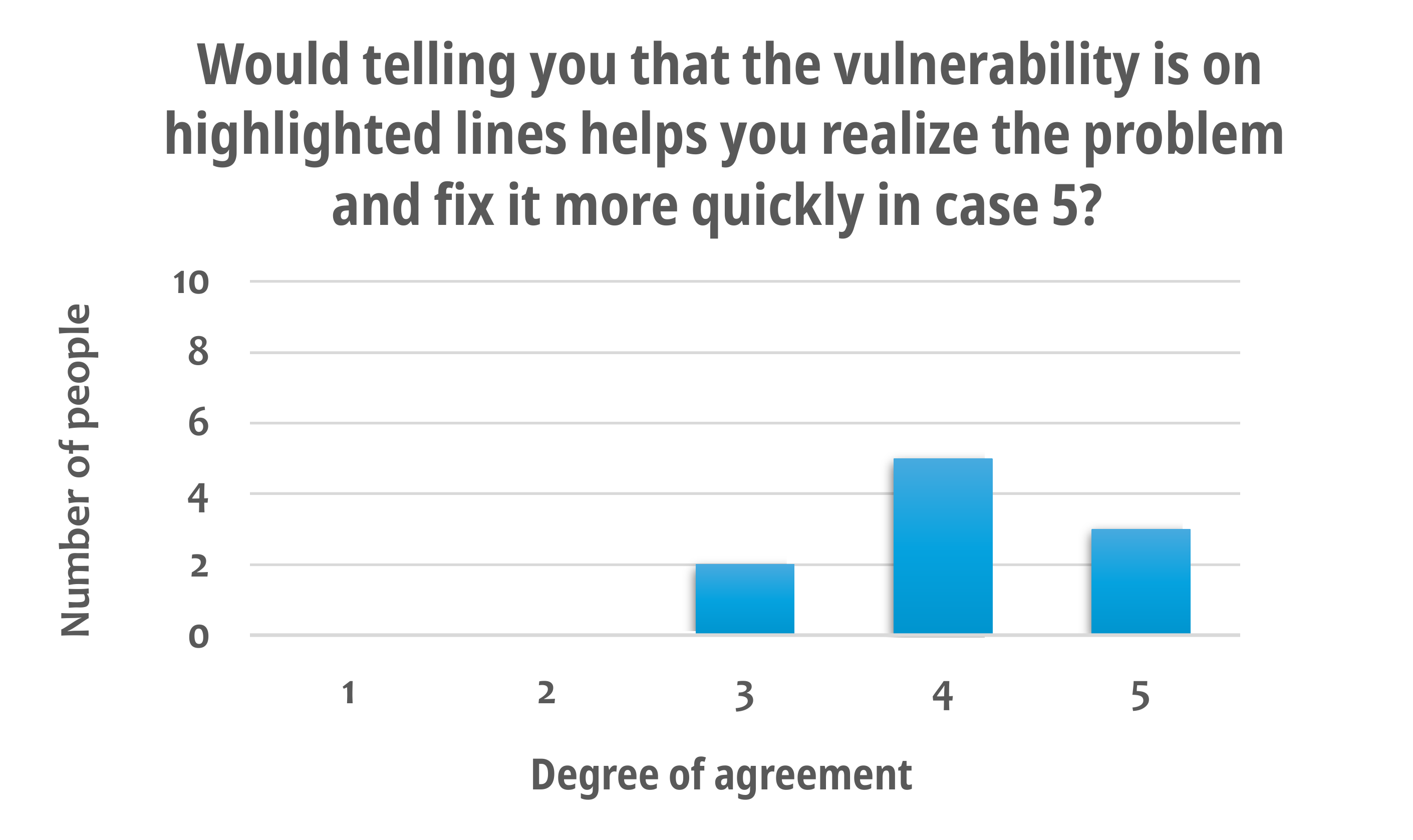}}
\caption{Results from the questionnaire about users' perspectives on vulnerability localization}
\label{fig:questionnaire}
\end{figure}

\subsubsection{Study on the effectiveness of vulnerability localization tools for developers in real-world scenarios}
To understand the effectiveness of vulnerability localization tools for developers in real-world scenarios, we conducted a study with 10 participants: 3 data analysts and 7 developers. Among them, three have 5 to 10 years of programming experience, five have 3 to 5 years, and the remaining two have 1 to 3 years. All participants frequently use Python; nine regularly use C/C++, three often use Java, and one regularly uses C\#. In the study, we first asked the participants some general questions about their experience with vulnerabilities. Then, we provided five specific vulnerability examples successfully localized by our tools and evaluated whether our tools helped users identify and fix the vulnerabilities in these examples.

Figure~\ref{fig:questionnaire} presents the responses of our participants to our questionnaire. Using a scale from 1 to 5, we measured participants’ agreement with our queries, where 1 indicated strong disagreement and 5 represented strong agreement. From Figure~\ref{fig:questionnaire} (a) and (b), we found that recognizing vulnerabilities is challenging for human developers. Most participants have limited confidence in discovering vulnerability issues independently. Even when informed that a given function is vulnerable, they still lack confidence in localizing and fixing the vulnerable statements. Figures~\ref{fig:questionnaire} (c) to (e) indicate that vulnerability detection tools assist users in identifying potential vulnerabilities and highlight a strong need for automatic vulnerability localization.

We randomly selected five vulnerable functions that can be correctly detected and localized by our proposed approach. We then asked participants if the localization information provided by our approach helped them recognize the vulnerability and fix it more quickly. According to the results shown in Figure~\ref{fig:questionnaire} (f) to (j), over 70\% of participants agreed that our proposed approach helped them identify and resolve the vulnerabilities in all the sample functions. Additionally, an average of 4.2 participants strongly agreed that our approach enabled them to address the problem more quickly in each sample vulnerable function, demonstrating the effectiveness of our proposed approach.

\begin{figure}[t]
\centering
\subfloat[\scriptsize Results from the questionnaire on the effectiveness of \tool in case 1]{
\includegraphics [width=4.8cm]{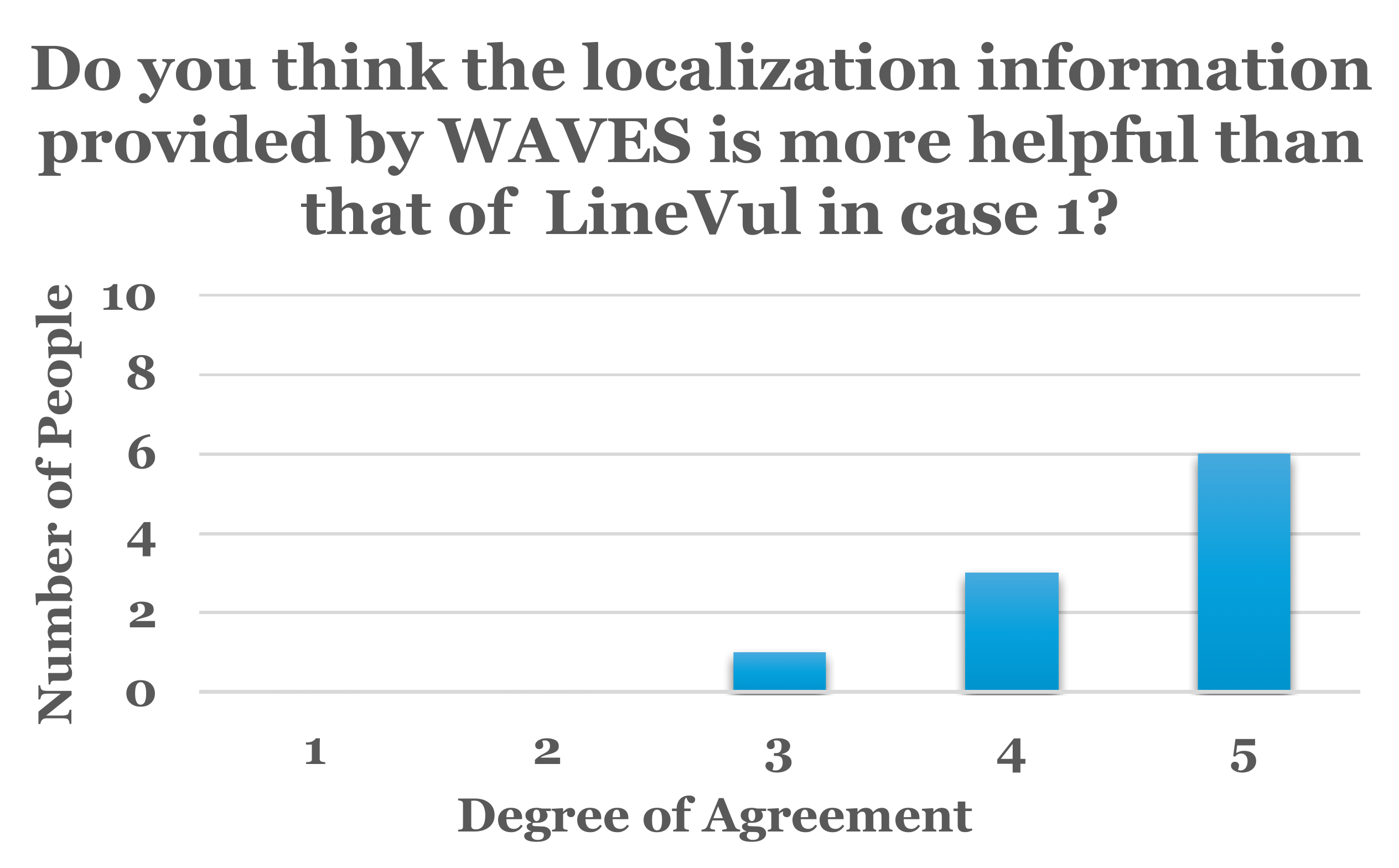}}
\hspace{0.1cm}
\subfloat[\scriptsize Results from the questionnaire on the effectiveness of \tool in case 2]{
\includegraphics [width=4.8cm]{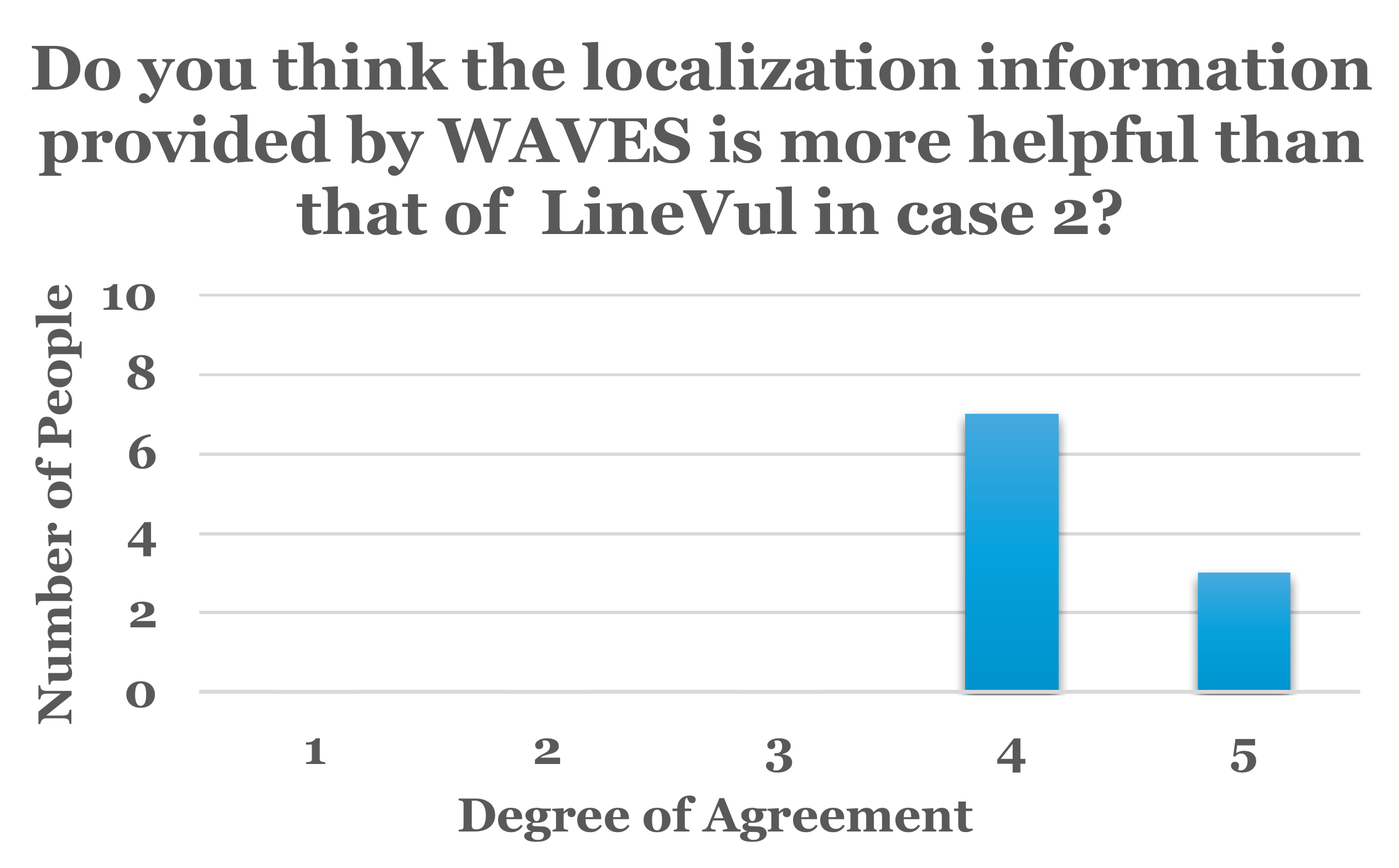}}
\hspace{0.1cm}
\subfloat[\scriptsize Results from the questionnaire on the effectiveness of \tool in case 3]{
\includegraphics [width=4.8cm]{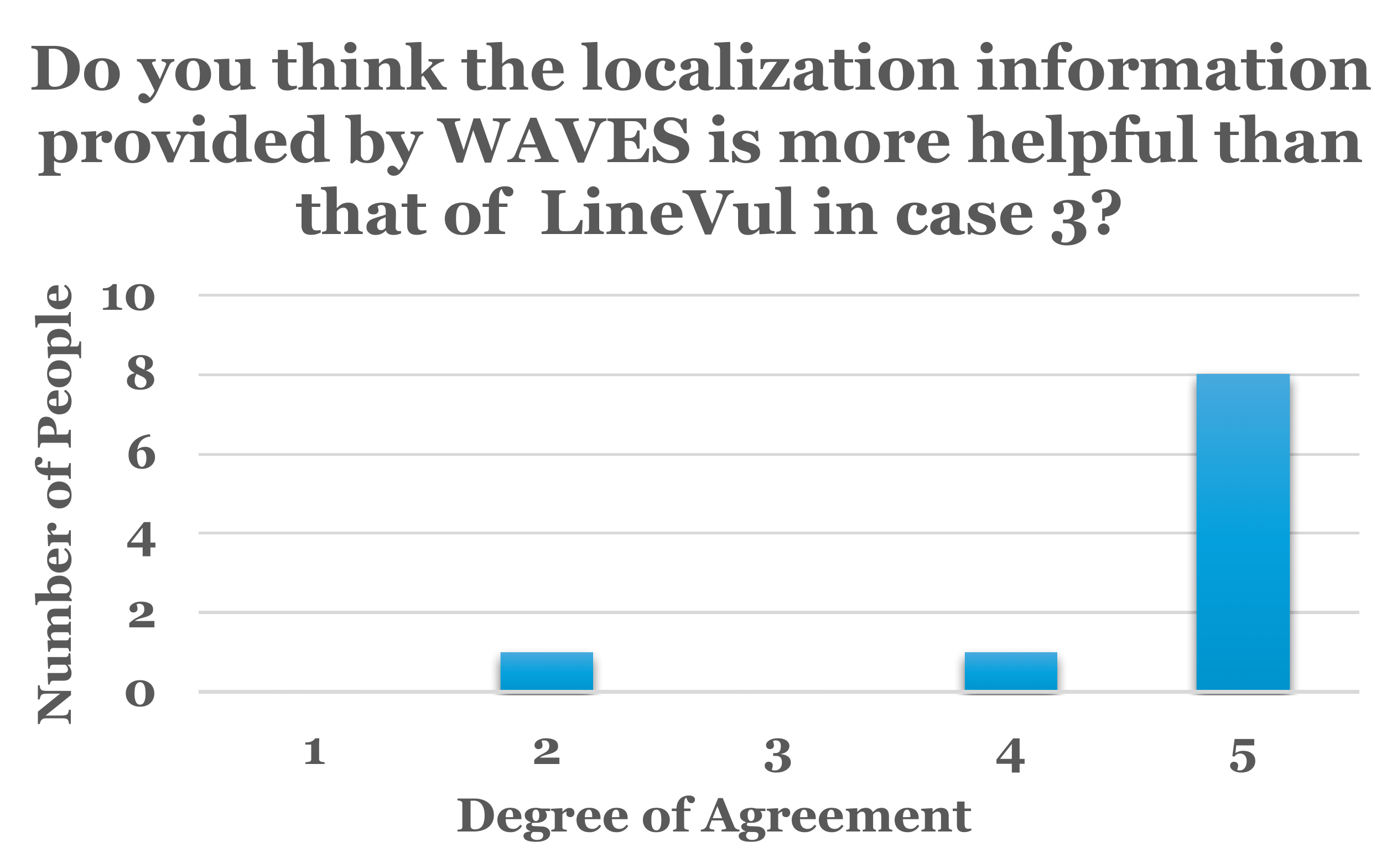}}
\quad
\subfloat[\scriptsize Results from the questionnaire on the effectiveness of \tool in case 4]{
\includegraphics [width=4.8cm]{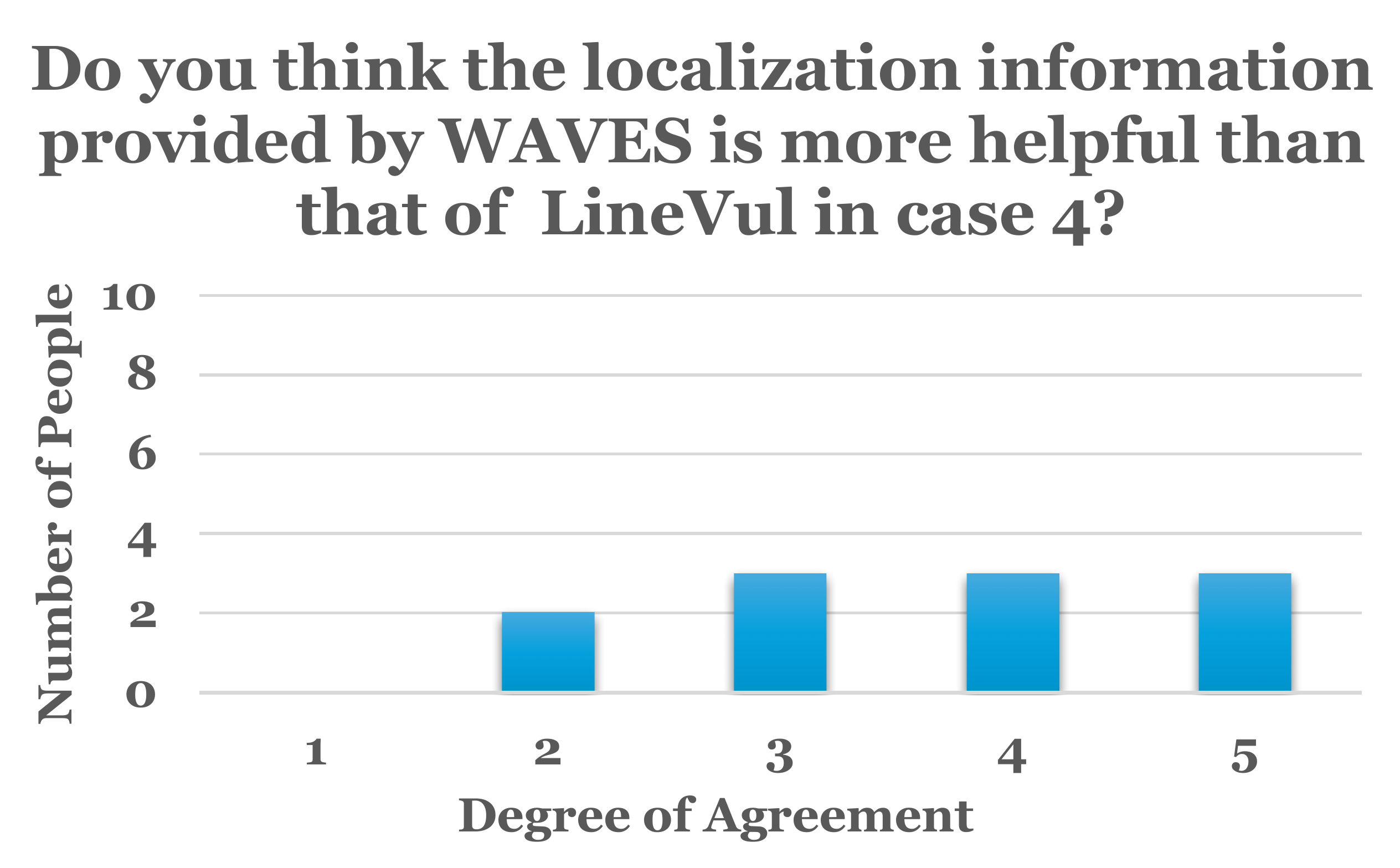}}
\hspace{0.1cm}
\subfloat[\scriptsize Results from the questionnaire on the effectiveness of \tool in case 5]{
\includegraphics [width=4.8cm]{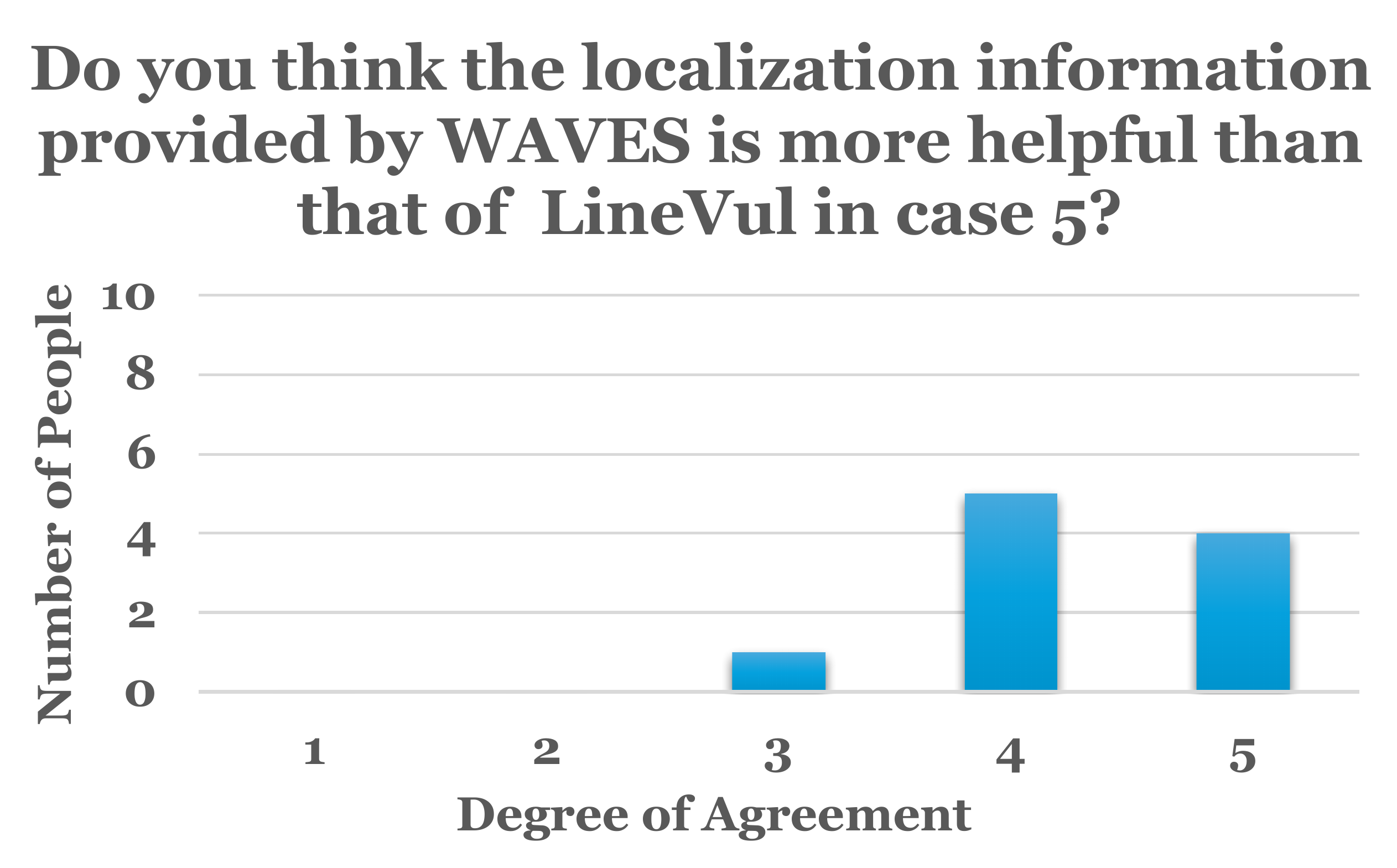}}
\caption{Results from the questionnaire about the effectiveness of \tool compared to LinVul on vulnerability localization}
\label{fig:questionnaire_compare}
\end{figure}

\subsubsection{Study on \tool's effectiveness compared to LineVul in vulnerability localization}
\revise{To evaluate the effectiveness of the vulnerability localization information provided by \tool compared to the baseline method LineVul, we conducted a study with 10 participants: 5 software developers, 4 software testing engineers, and 1 assistant professor in computer science. Among them, 2 participants have 3 to 5 years of programming experience, 4 have 5 to 10 years, and the remaining 4 have over 10 years of experience. All participants frequently use Python; 6 regularly work with C/C++, 7 often use Java, 2 frequently use JavaScript/TypeScript, and 1 regularly works with C\#.}

\revise{In Section~\ref{sec:rq1}, we have already demonstrated that \tool provides more accurate vulnerability localization information than LineVul. In this study, we aim to further validate whether these correct predictions from \tool actually help developers better recognize vulnerability issues compared to the incorrect predictions from LineVul. Specifically, we allow both \tool and LineVul to predict the most likely vulnerable statement and randomly select five test cases where \tool provides correct predictions while LineVul fails. Participants are then asked to evaluate how much the localization information provided by \tool improves their understanding of the vulnerability issue compared to LineVul, using a set of five predefined response options.}

\revise{Figure~\ref{fig:questionnaire_compare} presents the results of our user study. We observe that the localization information provided by \tool is generally more helpful than that from LineVul in assisting participants to identify vulnerability issues, demonstrating the effectiveness of \tool in vulnerability localization. However, its relative effectiveness varies across different test cases. Most participants agree that \tool's information is more useful in test cases 1, 2, 3, and 5, while in test case 4, half of the participants found its localization information to be as helpful as—or even less helpful than—that provided by LineVul, despite LineVul highlighting incorrect information. This discrepancy is likely due to the intrinsic difficulty of the vulnerability in test case 4; certain vulnerabilities are inherently hard to recognize, and even if the vulnerable statement is explicitly emphasized, participants may still fail to understand the issue.}

\section{Threats to Validity}
\label{sec:threats}
After careful analysis, we have identified several potential threats to the validity of our study:

\paragraph{Threats to External Validity}
Although we utilized three datasets for evaluation, we only assessed vulnerability localization performance on the dataset provided by Fan et al. The reason for not evaluating localization on the other two datasets is the lack of repaired data for vulnerable functions, which prevents us from labeling the vulnerable statements in these datasets. This challenge highlights the need for our proposed approach, as obtaining statement-level labels is often difficult in real-world scenarios. To effectively evaluate vulnerability localization, a dataset must include both the vulnerable code snippets and their corresponding fixes. While the D2A dataset~\cite{ZhengPLBEYLMS21} does contain such fixing information, Croft et al.~\cite{CroftBK23} reported that over 60\% of samples in D2A are mislabeled, rendering it unreliable. Therefore, we chose the dataset by Fan et al. as the testing dataset for vulnerability localization. While we have chosen three commonly used datasets to assess the effectiveness of our vulnerability detection approach, it is important to note that these datasets have limited size. Consequently, the results obtained from these datasets may not accurately reflect the performance of our approach in real-world scenarios.

\paragraph{Threats to Internal Validity}
In this paper, we adopt the hyperparameters from LineVul~\cite{linevul} to maintain consistency. While we acknowledge the potential impact of hyperparameters on the performance of our proposed \tool, we did not investigate their influence due to the considerable cost associated with model training. However, it is important to note that different hyperparameter settings may indeed affect \tool's performance. Among these settings, the maximum input token length holds particular significance. Currently, in \tool, we have set the maximum input token length to 512, discarding any additional tokens. The performance of \tool for longer code samples has not been thoroughly explored under this constraint.

Moreover, it is worth mentioning that certain vulnerabilities may be closely tied to the context of the code, such as use-after-free vulnerabilities. In some cases, the division of code segments could lead to the disappearance of existing vulnerabilities or even the emergence of new ones, potentially altering the label of the target function. Therefore, we need to carefully consider the implications of code segmentation on \tool's effectiveness in identifying such vulnerabilities. Further investigations into the impact of these factors are necessary for a comprehensive understanding of \tool's performance.

Furthermore, the vulnerability labels at the statement level are determined by checking whether the statements have been modified in the commit. Consequently, all the modified statements are considered vulnerable. However, simply altering a statement does not guarantee the presence of an actual vulnerability. This data pre-processing approach could potentially introduce biases into the vulnerable statement labels.


\section{Related Work}
\label{sec:suvery}
\subsection{Deep Learning-based Vulnerability Detection}
With the advent of deep learning technology, significant advancements have been made in various tasks, such as code retrieval and code generation. Consequently, researchers in the vulnerability detection field have also taken notice. The integration of deep learning into vulnerability detection approaches has resulted in a substantial improvement in performance compared to conventional methods. Existing studies in this area can be broadly categorized into token-based approaches and graph-based approaches, each utilizing different representations of the source code. In the following subsections, we provide a brief overview of these two types of approaches.

\subsubsection{Token-based Vulnerability Detection Approaches}

Several works \cite{vuldeepecker,dam2017automatic, sysevr} approach source code as flat sequences and adopt natural language processing techniques to represent input code and initialize tokenized code tokens with Word2Vec~\cite{wv}. Li et al.\cite{vuldeepecker} initiate the study of using deep learning for vulnerability detection. They transform programs into code gadgets consisting of multiple lines of code statements that are semantically related and propose VulDeePecker, a Bidirectional Long Short Time Memory (BLSTM) neural network with a dense layer to learn representations. To further represent programs into vectors that accommodate the syntax and semantic information suitable for vulnerability detection, \citet{sysevr} extract code slices according to data dependency and control dependency and also utilize BLSTM to obtain representation for detection. Another work \cite{dam2017automatic} leverages Convolutional Neural Networks (CNN) and Recurrent Neural Networks (RNNs) to directly learn features from source code. 

\subsubsection{Graph-based Vulnerability Detection Approaches} 

Source code is inherently structured and logical and has heterogeneous aspects of representation, such as Abstract Syntax Tree (AST), Control Flow Graph (CFG), Data Flow Graph (DFG), and Program Dependency Graph (PDG). Many works~\cite{devign, DeepWukong, ACGVD, reveal, IVDETECT, 9293321, vulcnn, ReGVD} represent source code into single code graphs or composite graphs to improve the syntactic and semantic information. \citet{devign} construct a heterogeneous joint graph consisting of AST, CFG, and DFG following~\cite{yamaguchi2014modeling}. They connect the neighboring leaf nodes of the AST to preserve the natural sequential order of the source code and utilize Gated Graph Neural Network (GGNN) with the convolution module for graph-level classification. \citet{reveal} extract the information of Code Property Graph (CPG) ~\cite{yamaguchi2014modeling} from the given function and adopts the technique of Word2Vec~\cite{wv} to initialize the embedding vector. They also utilize the GGNN model to learn graph representation and focus on solving the problem of dataset imbalance. \citet{IVDETECT} also extract multiple types of graphs, but unlike the above two methods which use GNN to directly learn the representation of joint graphs, they leverage Gate Recurrent Unit (GRU)~\cite{GRU} or Tree-LSTM~\cite{Tree-LSTM} to learn a single type of graph respectively and aggregate these vectors through convolution operation. \citet{DeepWukong} and \citet{vulcnn} both distill the function semantic information into a PDG which contains control-flow and data-flow details of source code, and they train a GNN model and a Convolutional Neural Network (CNN)~\cite{DBLP:conf/nips/KrizhevskySH12} model to detect the vulnerability, respectively.

\subsection{Deep Learning-based Statement-Level Vulnerability Detection and Localization}

Although deep learning-based approaches for function-level vulnerability detection have attracted a large number of researchers and achieved great progress, there are still limitations in practical applications. Even if the function-level vulnerability can be successfully detected, it still requires a large effort to locate the vulnerable statements inside the function if the vulnerable function contains many statements. 

\citet{IVDETECT} leveraged GCN for function-level predictions and GNNExplainer~\cite{GNNExplainer} to explore the subgraphs that contribute most to the predictions. However, this still cannot provide specific vulnerable statements. Many statement-level vulnerability detection approaches~\cite{linevul, linevd, velvet} have been proposed. \citet{velvet} and \citet{linevd} both provided labels for vulnerable statements in the training phase. \citet{velvet} propose an ensemble learning approach named VELVET which combines GGNN and Transformer~\cite{Transformer} to capture the local and global context of the source code. VELVET shows good performance on both vulnerability classification and localization. \citet{linevd} formulates statement-level vulnerability detection as a node classification task. They use CodeBERT~\cite{codebert} to initialize the embedding of each statement in the function and further leverage Graph Attention Network (GAT)~\cite{GAT} to update these statement embeddings according to the control and data dependency between statements.  The performance of these supervised approaches is sensitive to the quality and quantity of the annotated data. To solve this issue, \citet{linevul} propose an unsupervised approach name LineVul that leverages the attention mechanism of CodeBERT~\cite{codebert} to find the most likely vulnerable statements. LineVul uses CodeBERT to capture the long-term dependencies and semantic context of the statements and aggregate the attention score of each statement for statement-level vulnerability detection. The experiment results indicate that unsupervised statement-level vulnerability detection is feasible and has room for further improvement. 

\subsection{Multiple Instance Learning}

With the development of deep learning technology, people start to understand the importance of the data scale to deep learning. However, collecting fine-grained labeled data from the real world remains a big problem. The data we acquire from the real world are usually unlabelled and it will cost a lot of time and huge manpower to label these data. To alleviate the need for labeled data, the deep learning methods that require fewer data, including unsupervised learning and weakly supervised learning, have become a research hotspot. Multiple instance learning is a common method in weakly supervised learning. Under the framework of multiple instance learning, the training data are arranged into sets, which are called bags. Only the label of the bags will be provided and the label of every single data is not available during the training. The model with multiple instance learning can learn to predict either the instance label or the bag label by learning the labels of bags and distinguishing the positive instances and negative instances inside the bags. The feature of less demand for labeled data makes multiple instance learning widely used in different scenarios. We briefly introduce its application in a few areas.


In Computer Vision, Pinheiro et al.~\citep{PinheiroC15} convert the task of object segmentation to the task of inferring the pixels in the given image which belongs to the class of the object and propose a Convolutional Neural Network-based model with multiple instance learning framework for such a task. In the task of histopathology image analysis, the duration of pixel-level annotations is laborious and time-consuming. To address this problem, Chikontwe et al.~\citep{ChikontweKNGP20} propose a MIL-based approach that can jointly learn the embedding vectors at both bag-level and instance-level to diagnose cancer. In the task of retinal image classification, Tu et al.~\citep{abs-1906-04881} first attempt to combine the Graph Neural Networks with multiple instance learning and achieve state-of-art performance on the public datasets. Inspired by the dynamic routing in capsule networks, Yan et al.~\citep{YanWGFLH18} adopt a dynamic pooling function to replace the conventional max pooling or mean pooling function for multiple instance learning and achieve the state-of-art performance in the task of animal detection.

In the task of sentiment analysis for texts, Pappas et al.~\citep{PappasP14} encode each sentence or paragraph into a feature vector and adopt multiple instance regression (MIR) to assign importance weights to each of the sentences or paragraphs of a text to uncover their contribution to the aspect ratings. In the task of contextual advertisement, advertisers wish to avoid 
some specific content on web pages but it is difficult since most training pages are multi-topic and need people to label at the sub-document level. To address this challenge, Zhang et al.~\citep{ZhangSPN08} adopt multiple instance learning to detect and avoid content that is related to war, violence, and pornography or the negative opinion about their product. 

\section{Conclusion}
\label{sec:conclusion}
In this paper, we present a novel approach named \tool for vulnerability detection. \tool 
incorporates the multiple instance learning framework to predict whether a given function is vulnerable or not, while also offering precise localization information about the vulnerable statements within the function. Through experiments conducted on public datasets, we have demonstrated that \tool achieves comparable performance to previous baselines in function-level vulnerability detection, and outperforms state-of-the-art baselines in statement-level vulnerability localization.

In the future, we plan to investigate the dynamic assignment of distinct pseudo-labels at the statement level across different functions, aiming to further improve the accuracy of vulnerability localization. In addition, we will expand our dataset to include programs with longer code segments in order to systematically analyze how code length influences the performance of the proposed method.


\bibliographystyle{ACM-Reference-Format}
\bibliography{sample-base}


\begin{thebibliography}{52}


\ifx \showCODEN    \undefined \def \showCODEN     #1{\unskip}     \fi
\ifx \showDOI      \undefined \def \showDOI       #1{#1}\fi
\ifx \showISBNx    \undefined \def \showISBNx     #1{\unskip}     \fi
\ifx \showISBNxiii \undefined \def \showISBNxiii  #1{\unskip}     \fi
\ifx \showISSN     \undefined \def \showISSN      #1{\unskip}     \fi
\ifx \showLCCN     \undefined \def \showLCCN      #1{\unskip}     \fi
\ifx \shownote     \undefined \def \shownote      #1{#1}          \fi
\ifx \showarticletitle \undefined \def \showarticletitle #1{#1}   \fi
\ifx \showURL      \undefined \def \showURL       {\relax}        \fi
\providecommand\bibfield[2]{#2}
\providecommand\bibinfo[2]{#2}
\providecommand\natexlab[1]{#1}
\providecommand\showeprint[2][]{arXiv:#2}

\bibitem[CWE(0123)]%
        {CWE}
 \bibinfo{year}{20123}\natexlab{}.
\newblock \bibinfo{title}{2023 CWE Top 25 Most Dangerous Software Weaknesses}.
\newblock
  \bibinfo{howpublished}{\url{https://cwe.mitre.org/top25/archive/2023/2023_top25_list.html}}.
\newblock


\bibitem[Andrews et~al\mbox{.}(2002)]%
        {AndrewsTH02}
\bibfield{author}{\bibinfo{person}{Stuart Andrews}, \bibinfo{person}{Ioannis
  Tsochantaridis}, {and} \bibinfo{person}{Thomas Hofmann}.}
  \bibinfo{year}{2002}\natexlab{}.
\newblock \showarticletitle{Support Vector Machines for Multiple-Instance
  Learning}. In \bibinfo{booktitle}{\emph{Advances in Neural Information
  Processing Systems 15 [Neural Information Processing Systems, {NIPS} 2002,
  December 9-14, 2002, Vancouver, British Columbia, Canada]}},
  \bibfield{editor}{\bibinfo{person}{Suzanna Becker},
  \bibinfo{person}{Sebastian Thrun}, {and} \bibinfo{person}{Klaus Obermayer}}
  (Eds.). \bibinfo{publisher}{{MIT} Press}, \bibinfo{pages}{561--568}.
\newblock
\urldef\tempurl%
\url{https://proceedings.neurips.cc/paper/2002/hash/3e6260b81898beacda3d16db379ed329-Abstract.html}
\showURL{%
\tempurl}


\bibitem[Bergeron et~al\mbox{.}(2012)]%
        {BergeronMZBB12}
\bibfield{author}{\bibinfo{person}{Charles Bergeron},
  \bibinfo{person}{Gregory~M. Moore}, \bibinfo{person}{Jed Zaretzki},
  \bibinfo{person}{Curt~M. Breneman}, {and} \bibinfo{person}{Kristin~P.
  Bennett}.} \bibinfo{year}{2012}\natexlab{}.
\newblock \showarticletitle{Fast Bundle Algorithm for Multiple-Instance
  Learning}.
\newblock \bibinfo{journal}{\emph{{IEEE} Trans. Pattern Anal. Mach. Intell.}}
  \bibinfo{volume}{34}, \bibinfo{number}{6} (\bibinfo{year}{2012}),
  \bibinfo{pages}{1068--1079}.
\newblock
\urldef\tempurl%
\url{https://doi.org/10.1109/TPAMI.2011.194}
\showDOI{\tempurl}


\bibitem[Bhandari et~al\mbox{.}(2021)]%
        {BhandariNM21}
\bibfield{author}{\bibinfo{person}{Guru~Prasad Bhandari},
  \bibinfo{person}{Amara Naseer}, {and} \bibinfo{person}{Leon Moonen}.}
  \bibinfo{year}{2021}\natexlab{}.
\newblock \showarticletitle{CVEfixes: automated collection of vulnerabilities
  and their fixes from open-source software}. In
  \bibinfo{booktitle}{\emph{{PROMISE} '21: 17th International Conference on
  Predictive Models and Data Analytics in Software Engineering, Athens Greece,
  August 19-20, 2021}}, \bibfield{editor}{\bibinfo{person}{Shane McIntosh},
  \bibinfo{person}{Xin Xia}, {and} \bibinfo{person}{Sousuke Amasaki}} (Eds.).
  \bibinfo{publisher}{{ACM}}, \bibinfo{pages}{30--39}.
\newblock
\urldef\tempurl%
\url{https://doi.org/10.1145/3475960.3475985}
\showDOI{\tempurl}


\bibitem[Carbonneau et~al\mbox{.}(2018)]%
        {CarbonneauCGG18}
\bibfield{author}{\bibinfo{person}{Marc{-}Andr{\'{e}} Carbonneau},
  \bibinfo{person}{Veronika Cheplygina}, \bibinfo{person}{Eric Granger}, {and}
  \bibinfo{person}{Ghyslain Gagnon}.} \bibinfo{year}{2018}\natexlab{}.
\newblock \showarticletitle{Multiple instance learning: {A} survey of problem
  characteristics and applications}.
\newblock \bibinfo{journal}{\emph{Pattern Recognit.}}  \bibinfo{volume}{77}
  (\bibinfo{year}{2018}), \bibinfo{pages}{329--353}.
\newblock
\urldef\tempurl%
\url{https://doi.org/10.1016/J.PATCOG.2017.10.009}
\showDOI{\tempurl}


\bibitem[Chakraborty et~al\mbox{.}(2022)]%
        {reveal}
\bibfield{author}{\bibinfo{person}{Saikat Chakraborty}, \bibinfo{person}{Rahul
  Krishna}, \bibinfo{person}{Yangruibo Ding}, {and} \bibinfo{person}{Baishakhi
  Ray}.} \bibinfo{year}{2022}\natexlab{}.
\newblock \showarticletitle{Deep Learning Based Vulnerability Detection: Are We
  There Yet?}
\newblock \bibinfo{journal}{\emph{{IEEE} Trans. Software Eng.}}
  \bibinfo{volume}{48}, \bibinfo{number}{9} (\bibinfo{year}{2022}),
  \bibinfo{pages}{3280--3296}.
\newblock


\bibitem[Cheng et~al\mbox{.}(2021)]%
        {DeepWukong}
\bibfield{author}{\bibinfo{person}{Xiao Cheng}, \bibinfo{person}{Haoyu Wang},
  \bibinfo{person}{Jiayi Hua}, \bibinfo{person}{Guoai Xu}, {and}
  \bibinfo{person}{Yulei Sui}.} \bibinfo{year}{2021}\natexlab{}.
\newblock \showarticletitle{DeepWukong: Statically Detecting Software
  Vulnerabilities Using Deep Graph Neural Network}.
\newblock \bibinfo{journal}{\emph{{ACM} Trans. Softw. Eng. Methodol.}}
  \bibinfo{volume}{30}, \bibinfo{number}{3} (\bibinfo{year}{2021}),
  \bibinfo{pages}{38:1--38:33}.
\newblock


\bibitem[Chikontwe et~al\mbox{.}(2020)]%
        {ChikontweKNGP20}
\bibfield{author}{\bibinfo{person}{Philip Chikontwe}, \bibinfo{person}{Meejeong
  Kim}, \bibinfo{person}{Soo~Jeong Nam}, \bibinfo{person}{Heounjeong Go}, {and}
  \bibinfo{person}{Sang~Hyun Park}.} \bibinfo{year}{2020}\natexlab{}.
\newblock \showarticletitle{Multiple Instance Learning with Center Embeddings
  for Histopathology Classification}. In \bibinfo{booktitle}{\emph{Medical
  Image Computing and Computer Assisted Intervention - {MICCAI} 2020 - 23rd
  International Conference, Lima, Peru, October 4-8, 2020, Proceedings, Part
  {V}}} \emph{(\bibinfo{series}{Lecture Notes in Computer Science},
  Vol.~\bibinfo{volume}{12265})}, \bibfield{editor}{\bibinfo{person}{Anne~L.
  Martel}, \bibinfo{person}{Purang Abolmaesumi}, \bibinfo{person}{Danail
  Stoyanov}, \bibinfo{person}{Diana Mateus}, \bibinfo{person}{Maria~A.
  Zuluaga}, \bibinfo{person}{S.~Kevin Zhou}, \bibinfo{person}{Daniel
  Racoceanu}, {and} \bibinfo{person}{Leo Joskowicz}} (Eds.).
  \bibinfo{publisher}{Springer}, \bibinfo{pages}{519--528}.
\newblock
\urldef\tempurl%
\url{https://doi.org/10.1007/978-3-030-59722-1\_50}
\showDOI{\tempurl}


\bibitem[Chung et~al\mbox{.}(2014)]%
        {GRU}
\bibfield{author}{\bibinfo{person}{Junyoung Chung},
  \bibinfo{person}{{\c{C}}aglar G{\"{u}}l{\c{c}}ehre},
  \bibinfo{person}{KyungHyun Cho}, {and} \bibinfo{person}{Yoshua Bengio}.}
  \bibinfo{year}{2014}\natexlab{}.
\newblock \showarticletitle{Empirical Evaluation of Gated Recurrent Neural
  Networks on Sequence Modeling}.
\newblock \bibinfo{journal}{\emph{CoRR}}  \bibinfo{volume}{abs/1412.3555}
  (\bibinfo{year}{2014}).
\newblock


\bibitem[Croft et~al\mbox{.}(2023)]%
        {CroftBK23}
\bibfield{author}{\bibinfo{person}{Roland Croft}, \bibinfo{person}{Muhammad~Ali
  Babar}, {and} \bibinfo{person}{M.~Mehdi Kholoosi}.}
  \bibinfo{year}{2023}\natexlab{}.
\newblock \showarticletitle{Data Quality for Software Vulnerability Datasets}.
  In \bibinfo{booktitle}{\emph{45th {IEEE/ACM} International Conference on
  Software Engineering, {ICSE} 2023, Melbourne, Australia, May 14-20, 2023}}.
  \bibinfo{publisher}{{IEEE}}, \bibinfo{pages}{121--133}.
\newblock
\urldef\tempurl%
\url{https://doi.org/10.1109/ICSE48619.2023.00022}
\showDOI{\tempurl}


\bibitem[Dam et~al\mbox{.}(2017)]%
        {dam2017automatic}
\bibfield{author}{\bibinfo{person}{Hoa~Khanh Dam}, \bibinfo{person}{Truyen
  Tran}, \bibinfo{person}{Trang Pham}, \bibinfo{person}{Shien~Wee Ng},
  \bibinfo{person}{John Grundy}, {and} \bibinfo{person}{Aditya Ghose}.}
  \bibinfo{year}{2017}\natexlab{}.
\newblock \showarticletitle{Automatic feature learning for vulnerability
  prediction}.
\newblock \bibinfo{journal}{\emph{CoRR}}  \bibinfo{volume}{abs/1708.02368}
  (\bibinfo{year}{2017}).
\newblock
\showeprint[arXiv]{1708.02368}
\urldef\tempurl%
\url{http://arxiv.org/abs/1708.02368}
\showURL{%
\tempurl}


\bibitem[Ding et~al\mbox{.}(2022)]%
        {velvet}
\bibfield{author}{\bibinfo{person}{Yangruibo Ding}, \bibinfo{person}{Sahil
  Suneja}, \bibinfo{person}{Yunhui Zheng}, \bibinfo{person}{Jim Laredo},
  \bibinfo{person}{Alessandro Morari}, \bibinfo{person}{Gail~E. Kaiser}, {and}
  \bibinfo{person}{Baishakhi Ray}.} \bibinfo{year}{2022}\natexlab{}.
\newblock \showarticletitle{{VELVET:} a noVel Ensemble Learning approach to
  automatically locate VulnErable sTatements}. In
  \bibinfo{booktitle}{\emph{{IEEE} International Conference on Software
  Analysis, Evolution and Reengineering, {SANER} 2022, Honolulu, HI, USA, March
  15-18, 2022}}. \bibinfo{publisher}{{IEEE}}, \bibinfo{pages}{959--970}.
\newblock


\bibitem[Facebook(2021)]%
        {Infer}
\bibfield{author}{\bibinfo{person}{Facebook}.} \bibinfo{year}{2021}\natexlab{}.
\newblock \bibinfo{title}{Infer}.
\newblock \bibinfo{howpublished}{\url{https://fbinfer.com/.}}.
\newblock


\bibitem[Fan et~al\mbox{.}(2020)]%
        {fan}
\bibfield{author}{\bibinfo{person}{Jiahao Fan}, \bibinfo{person}{Yi Li},
  \bibinfo{person}{Shaohua Wang}, {and} \bibinfo{person}{Tien~N. Nguyen}.}
  \bibinfo{year}{2020}\natexlab{}.
\newblock \showarticletitle{A {C/C++} Code Vulnerability Dataset with Code
  Changes and {CVE} Summaries}. In \bibinfo{booktitle}{\emph{{MSR} '20: 17th
  International Conference on Mining Software Repositories, Seoul, Republic of
  Korea, 29-30 June, 2020}}. \bibinfo{publisher}{{ACM}},
  \bibinfo{pages}{508--512}.
\newblock


\bibitem[Feng et~al\mbox{.}(2020)]%
        {codebert}
\bibfield{author}{\bibinfo{person}{Zhangyin Feng}, \bibinfo{person}{Daya Guo},
  \bibinfo{person}{Duyu Tang}, \bibinfo{person}{Nan Duan},
  \bibinfo{person}{Xiaocheng Feng}, \bibinfo{person}{Ming Gong},
  \bibinfo{person}{Linjun Shou}, \bibinfo{person}{Bing Qin},
  \bibinfo{person}{Ting Liu}, \bibinfo{person}{Daxin Jiang}, {and}
  \bibinfo{person}{Ming Zhou}.} \bibinfo{year}{2020}\natexlab{}.
\newblock \showarticletitle{CodeBERT: {A} Pre-Trained Model for Programming and
  Natural Languages}. In \bibinfo{booktitle}{\emph{Findings of the Association
  for Computational Linguistics: {EMNLP} 2020, Online Event, 16-20 November
  2020}} \emph{(\bibinfo{series}{Findings of {ACL}},
  Vol.~\bibinfo{volume}{{EMNLP} 2020})},
  \bibfield{editor}{\bibinfo{person}{Trevor Cohn}, \bibinfo{person}{Yulan He},
  {and} \bibinfo{person}{Yang Liu}} (Eds.). \bibinfo{publisher}{Association for
  Computational Linguistics}, \bibinfo{pages}{1536--1547}.
\newblock


\bibitem[Fu and Tantithamthavorn(2022)]%
        {linevul}
\bibfield{author}{\bibinfo{person}{Michael Fu} {and} \bibinfo{person}{Chakkrit
  Tantithamthavorn}.} \bibinfo{year}{2022}\natexlab{}.
\newblock \showarticletitle{LineVul: {A} Transformer-based Line-Level
  Vulnerability Prediction}. In \bibinfo{booktitle}{\emph{19th {IEEE/ACM}
  International Conference on Mining Software Repositories, {MSR} 2022,
  Pittsburgh, PA, USA, May 23-24, 2022}}. \bibinfo{publisher}{{ACM}},
  \bibinfo{pages}{608--620}.
\newblock


\bibitem[Goodin({[n.\,d.]})]%
        {Goodin}
\bibfield{author}{\bibinfo{person}{Dan Goodin}.}
  \bibinfo{year}{[n.\,d.]}\natexlab{}.
\newblock \showarticletitle{An NSA-derived ransomware worm is shutting down
  computers worldwide (2017)}.
\newblock
\urldef\tempurl%
\url{https://arstechnica.com/ information-technology/2017/05/}
\showURL{%
\tempurl}


\bibitem[Guo et~al\mbox{.}(2022)]%
        {GuoLDW0022}
\bibfield{author}{\bibinfo{person}{Daya Guo}, \bibinfo{person}{Shuai Lu},
  \bibinfo{person}{Nan Duan}, \bibinfo{person}{Yanlin Wang},
  \bibinfo{person}{Ming Zhou}, {and} \bibinfo{person}{Jian Yin}.}
  \bibinfo{year}{2022}\natexlab{}.
\newblock \showarticletitle{UniXcoder: Unified Cross-Modal Pre-training for
  Code Representation}. In \bibinfo{booktitle}{\emph{Proceedings of the 60th
  Annual Meeting of the Association for Computational Linguistics (Volume 1:
  Long Papers), {ACL} 2022, Dublin, Ireland, May 22-27, 2022}},
  \bibfield{editor}{\bibinfo{person}{Smaranda Muresan},
  \bibinfo{person}{Preslav Nakov}, {and} \bibinfo{person}{Aline Villavicencio}}
  (Eds.). \bibinfo{publisher}{Association for Computational Linguistics},
  \bibinfo{pages}{7212--7225}.
\newblock
\urldef\tempurl%
\url{https://doi.org/10.18653/V1/2022.ACL-LONG.499}
\showDOI{\tempurl}


\bibitem[Guo et~al\mbox{.}(2021)]%
        {GuoRLFT0ZDSFTDC21}
\bibfield{author}{\bibinfo{person}{Daya Guo}, \bibinfo{person}{Shuo Ren},
  \bibinfo{person}{Shuai Lu}, \bibinfo{person}{Zhangyin Feng},
  \bibinfo{person}{Duyu Tang}, \bibinfo{person}{Shujie Liu},
  \bibinfo{person}{Long Zhou}, \bibinfo{person}{Nan Duan},
  \bibinfo{person}{Alexey Svyatkovskiy}, \bibinfo{person}{Shengyu Fu},
  \bibinfo{person}{Michele Tufano}, \bibinfo{person}{Shao~Kun Deng},
  \bibinfo{person}{Colin~B. Clement}, \bibinfo{person}{Dawn Drain},
  \bibinfo{person}{Neel Sundaresan}, \bibinfo{person}{Jian Yin},
  \bibinfo{person}{Daxin Jiang}, {and} \bibinfo{person}{Ming Zhou}.}
  \bibinfo{year}{2021}\natexlab{}.
\newblock \showarticletitle{GraphCodeBERT: Pre-training Code Representations
  with Data Flow}. In \bibinfo{booktitle}{\emph{9th International Conference on
  Learning Representations, {ICLR} 2021, Virtual Event, Austria, May 3-7,
  2021}}. \bibinfo{publisher}{OpenReview.net}.
\newblock
\urldef\tempurl%
\url{https://openreview.net/forum?id=jLoC4ez43PZ}
\showURL{%
\tempurl}


\bibitem[Hin et~al\mbox{.}(2022)]%
        {linevd}
\bibfield{author}{\bibinfo{person}{David Hin}, \bibinfo{person}{Andrey Kan},
  \bibinfo{person}{Huaming Chen}, {and} \bibinfo{person}{Muhammad~Ali Babar}.}
  \bibinfo{year}{2022}\natexlab{}.
\newblock \showarticletitle{LineVD: Statement-level Vulnerability Detection
  using Graph Neural Networks}. In \bibinfo{booktitle}{\emph{19th {IEEE/ACM}
  International Conference on Mining Software Repositories, {MSR} 2022,
  Pittsburgh, PA, USA, May 23-24, 2022}}. \bibinfo{publisher}{{ACM}},
  \bibinfo{pages}{596--607}.
\newblock


\bibitem[Hochreiter and Schmidhuber(1997)]%
        {LSTM}
\bibfield{author}{\bibinfo{person}{Sepp Hochreiter} {and}
  \bibinfo{person}{J{\"{u}}rgen Schmidhuber}.} \bibinfo{year}{1997}\natexlab{}.
\newblock \showarticletitle{Long Short-Term Memory}.
\newblock \bibinfo{journal}{\emph{Neural Comput.}} \bibinfo{volume}{9},
  \bibinfo{number}{8} (\bibinfo{year}{1997}), \bibinfo{pages}{1735--1780}.
\newblock


\bibitem[Israel(2021)]%
        {Checkmarx}
\bibfield{author}{\bibinfo{person}{Israel}.} \bibinfo{year}{2021}\natexlab{}.
\newblock \bibinfo{title}{Checkmarx}.
\newblock \bibinfo{howpublished}{\url{https://checkmarx.com/.}}.
\newblock


\bibitem[Kotzias et~al\mbox{.}(2015)]%
        {DBLP:conf/kdd/KotziasDFS15}
\bibfield{author}{\bibinfo{person}{Dimitrios Kotzias}, \bibinfo{person}{Misha
  Denil}, \bibinfo{person}{Nando de Freitas}, {and} \bibinfo{person}{Padhraic
  Smyth}.} \bibinfo{year}{2015}\natexlab{}.
\newblock \showarticletitle{From Group to Individual Labels Using Deep
  Features}. In \bibinfo{booktitle}{\emph{Proceedings of the 21th {ACM}
  {SIGKDD} International Conference on Knowledge Discovery and Data Mining,
  Sydney, NSW, Australia, August 10-13, 2015}},
  \bibfield{editor}{\bibinfo{person}{Longbing Cao}, \bibinfo{person}{Chengqi
  Zhang}, \bibinfo{person}{Thorsten Joachims}, \bibinfo{person}{Geoffrey~I.
  Webb}, \bibinfo{person}{Dragos~D. Margineantu}, {and} \bibinfo{person}{Graham
  Williams}} (Eds.). \bibinfo{publisher}{{ACM}}, \bibinfo{pages}{597--606}.
\newblock


\bibitem[Krizhevsky et~al\mbox{.}(2012)]%
        {DBLP:conf/nips/KrizhevskySH12}
\bibfield{author}{\bibinfo{person}{Alex Krizhevsky}, \bibinfo{person}{Ilya
  Sutskever}, {and} \bibinfo{person}{Geoffrey~E. Hinton}.}
  \bibinfo{year}{2012}\natexlab{}.
\newblock \showarticletitle{ImageNet Classification with Deep Convolutional
  Neural Networks}. In \bibinfo{booktitle}{\emph{Advances in Neural Information
  Processing Systems 25: 26th Annual Conference on Neural Information
  Processing Systems 2012. Proceedings of a meeting held December 3-6, 2012,
  Lake Tahoe, Nevada, United States}},
  \bibfield{editor}{\bibinfo{person}{Peter~L. Bartlett},
  \bibinfo{person}{Fernando C.~N. Pereira}, \bibinfo{person}{Christopher J.~C.
  Burges}, \bibinfo{person}{L{\'{e}}on Bottou}, {and}
  \bibinfo{person}{Kilian~Q. Weinberger}} (Eds.). \bibinfo{pages}{1106--1114}.
\newblock


\bibitem[Li et~al\mbox{.}(2021a)]%
        {ACGVD}
\bibfield{author}{\bibinfo{person}{Min Li}, \bibinfo{person}{Chunfang Li},
  \bibinfo{person}{Shuailou Li}, \bibinfo{person}{Yanna Wu},
  \bibinfo{person}{Boyang Zhang}, {and} \bibinfo{person}{Yu Wen}.}
  \bibinfo{year}{2021}\natexlab{a}.
\newblock \showarticletitle{{ACGVD:} Vulnerability Detection Based on
  Comprehensive Graph via Graph Neural Network with Attention}. In
  \bibinfo{booktitle}{\emph{Information and Communications Security - 23rd
  International Conference, {ICICS} 2021, Chongqing, China, November 19-21,
  2021, Proceedings, Part {I}}} \emph{(\bibinfo{series}{Lecture Notes in
  Computer Science}, Vol.~\bibinfo{volume}{12918})},
  \bibfield{editor}{\bibinfo{person}{Debin Gao}, \bibinfo{person}{Qi~Li},
  \bibinfo{person}{Xiaohong Guan}, {and} \bibinfo{person}{Xiaofeng Liao}}
  (Eds.). \bibinfo{publisher}{Springer}, \bibinfo{pages}{243--259}.
\newblock


\bibitem[Li et~al\mbox{.}(2021b)]%
        {IVDETECT}
\bibfield{author}{\bibinfo{person}{Yi Li}, \bibinfo{person}{Shaohua Wang},
  {and} \bibinfo{person}{Tien~N. Nguyen}.} \bibinfo{year}{2021}\natexlab{b}.
\newblock \showarticletitle{Vulnerability detection with fine-grained
  interpretations}. In \bibinfo{booktitle}{\emph{{ESEC/FSE} '21: 29th {ACM}
  Joint European Software Engineering Conference and Symposium on the
  Foundations of Software Engineering, Athens, Greece, August 23-28, 2021}}.
  \bibinfo{publisher}{{ACM}}, \bibinfo{pages}{292--303}.
\newblock


\bibitem[Li et~al\mbox{.}(2022)]%
        {sysevr}
\bibfield{author}{\bibinfo{person}{Zhen Li}, \bibinfo{person}{Deqing Zou},
  \bibinfo{person}{Shouhuai Xu}, \bibinfo{person}{Hai Jin},
  \bibinfo{person}{Yawei Zhu}, {and} \bibinfo{person}{Zhaoxuan Chen}.}
  \bibinfo{year}{2022}\natexlab{}.
\newblock \showarticletitle{SySeVR: {A} Framework for Using Deep Learning to
  Detect Software Vulnerabilities}.
\newblock \bibinfo{journal}{\emph{{IEEE} Trans. Dependable Secur. Comput.}}
  \bibinfo{volume}{19}, \bibinfo{number}{4} (\bibinfo{year}{2022}),
  \bibinfo{pages}{2244--2258}.
\newblock


\bibitem[Li et~al\mbox{.}(2018)]%
        {vuldeepecker}
\bibfield{author}{\bibinfo{person}{Zhen Li}, \bibinfo{person}{Deqing Zou},
  \bibinfo{person}{Shouhuai Xu}, \bibinfo{person}{Xinyu Ou},
  \bibinfo{person}{Hai Jin}, \bibinfo{person}{Sujuan Wang},
  \bibinfo{person}{Zhijun Deng}, {and} \bibinfo{person}{Yuyi Zhong}.}
  \bibinfo{year}{2018}\natexlab{}.
\newblock \showarticletitle{Vuldeepecker: A deep learning-based system for
  vulnerability detection}.
\newblock \bibinfo{journal}{\emph{arXiv preprint arXiv:1801.01681}}
  (\bibinfo{year}{2018}).
\newblock


\bibitem[Lin et~al\mbox{.}(2020)]%
        {LinGGHD20}
\bibfield{author}{\bibinfo{person}{Tsung{-}Yi Lin}, \bibinfo{person}{Priya
  Goyal}, \bibinfo{person}{Ross~B. Girshick}, \bibinfo{person}{Kaiming He},
  {and} \bibinfo{person}{Piotr Doll{\'{a}}r}.} \bibinfo{year}{2020}\natexlab{}.
\newblock \showarticletitle{Focal Loss for Dense Object Detection}.
\newblock \bibinfo{journal}{\emph{{IEEE} Trans. Pattern Anal. Mach. Intell.}}
  \bibinfo{volume}{42}, \bibinfo{number}{2} (\bibinfo{year}{2020}),
  \bibinfo{pages}{318--327}.
\newblock
\urldef\tempurl%
\url{https://doi.org/10.1109/TPAMI.2018.2858826}
\showDOI{\tempurl}


\bibitem[Loshchilov and Hutter(2017)]%
        {adamw}
\bibfield{author}{\bibinfo{person}{Ilya Loshchilov} {and}
  \bibinfo{person}{Frank Hutter}.} \bibinfo{year}{2017}\natexlab{}.
\newblock \showarticletitle{Fixing Weight Decay Regularization in Adam}.
\newblock \bibinfo{journal}{\emph{CoRR}}  \bibinfo{volume}{abs/1711.05101}
  (\bibinfo{year}{2017}).
\newblock


\bibitem[Mikolov et~al\mbox{.}(2013)]%
        {wv}
\bibfield{author}{\bibinfo{person}{Tom{\'{a}}s Mikolov}, \bibinfo{person}{Kai
  Chen}, \bibinfo{person}{Greg Corrado}, {and} \bibinfo{person}{Jeffrey Dean}.}
  \bibinfo{year}{2013}\natexlab{}.
\newblock \showarticletitle{Efficient Estimation of Word Representations in
  Vector Space}. In \bibinfo{booktitle}{\emph{1st International Conference on
  Learning Representations, {ICLR} 2013}}.
\newblock


\bibitem[Nguyen et~al\mbox{.}(2022)]%
        {ReGVD}
\bibfield{author}{\bibinfo{person}{Van{-}Anh Nguyen}, \bibinfo{person}{Dai~Quoc
  Nguyen}, \bibinfo{person}{Van Nguyen}, \bibinfo{person}{Trung Le},
  \bibinfo{person}{Quan~Hung Tran}, {and} \bibinfo{person}{Dinh Phung}.}
  \bibinfo{year}{2022}\natexlab{}.
\newblock \showarticletitle{ReGVD: Revisiting Graph Neural Networks for
  Vulnerability Detection}. In \bibinfo{booktitle}{\emph{44th {IEEE/ACM}
  International Conference on Software Engineering: Companion Proceedings,
  {ICSE} Companion 2022, Pittsburgh, PA, USA, May 22-24, 2022}}.
  \bibinfo{publisher}{{ACM/IEEE}}, \bibinfo{pages}{178--182}.
\newblock


\bibitem[Pappas and Popescu{-}Belis(2014)]%
        {PappasP14}
\bibfield{author}{\bibinfo{person}{Nikolaos Pappas} {and}
  \bibinfo{person}{Andrei Popescu{-}Belis}.} \bibinfo{year}{2014}\natexlab{}.
\newblock \showarticletitle{Explaining the Stars: Weighted Multiple-Instance
  Learning for Aspect-Based Sentiment Analysis}. In
  \bibinfo{booktitle}{\emph{Proceedings of the 2014 Conference on Empirical
  Methods in Natural Language Processing, {EMNLP} 2014, October 25-29, 2014,
  Doha, Qatar, {A} meeting of SIGDAT, a Special Interest Group of the {ACL}}},
  \bibfield{editor}{\bibinfo{person}{Alessandro Moschitti},
  \bibinfo{person}{Bo~Pang}, {and} \bibinfo{person}{Walter Daelemans}} (Eds.).
  \bibinfo{publisher}{{ACL}}, \bibinfo{pages}{455--466}.
\newblock
\urldef\tempurl%
\url{https://doi.org/10.3115/v1/d14-1052}
\showDOI{\tempurl}


\bibitem[Pinheiro and Collobert(2015)]%
        {PinheiroC15}
\bibfield{author}{\bibinfo{person}{Pedro H.~O. Pinheiro} {and}
  \bibinfo{person}{Ronan Collobert}.} \bibinfo{year}{2015}\natexlab{}.
\newblock \showarticletitle{From image-level to pixel-level labeling with
  Convolutional Networks}. In \bibinfo{booktitle}{\emph{{IEEE} Conference on
  Computer Vision and Pattern Recognition, {CVPR} 2015, Boston, MA, USA, June
  7-12, 2015}}. \bibinfo{publisher}{{IEEE} Computer Society},
  \bibinfo{pages}{1713--1721}.
\newblock
\urldef\tempurl%
\url{https://doi.org/10.1109/CVPR.2015.7298780}
\showDOI{\tempurl}


\bibitem[Rahmani and Goldman(2006)]%
        {RahmaniG06}
\bibfield{author}{\bibinfo{person}{Rouhollah Rahmani} {and}
  \bibinfo{person}{Sally~A. Goldman}.} \bibinfo{year}{2006}\natexlab{}.
\newblock \showarticletitle{{MISSL:} multiple-instance semi-supervised
  learning}. In \bibinfo{booktitle}{\emph{Machine Learning, Proceedings of the
  Twenty-Third International Conference {(ICML} 2006), Pittsburgh,
  Pennsylvania, USA, June 25-29, 2006}} \emph{(\bibinfo{series}{{ACM}
  International Conference Proceeding Series}, Vol.~\bibinfo{volume}{148})},
  \bibfield{editor}{\bibinfo{person}{William~W. Cohen} {and}
  \bibinfo{person}{Andrew~W. Moore}} (Eds.). \bibinfo{publisher}{{ACM}},
  \bibinfo{pages}{705--712}.
\newblock
\urldef\tempurl%
\url{https://doi.org/10.1145/1143844.1143933}
\showDOI{\tempurl}


\bibitem[Russell et~al\mbox{.}(2018)]%
        {Russell}
\bibfield{author}{\bibinfo{person}{Rebecca~L. Russell},
  \bibinfo{person}{Louis~Y. Kim}, \bibinfo{person}{Lei~H. Hamilton},
  \bibinfo{person}{Tomo Lazovich}, \bibinfo{person}{Jacob Harer},
  \bibinfo{person}{Onur Ozdemir}, \bibinfo{person}{Paul~M. Ellingwood}, {and}
  \bibinfo{person}{Marc~W. McConley}.} \bibinfo{year}{2018}\natexlab{}.
\newblock \showarticletitle{Automated Vulnerability Detection in Source Code
  Using Deep Representation Learning}. In \bibinfo{booktitle}{\emph{17th {IEEE}
  International Conference on Machine Learning and Applications, {ICMLA} 2018,
  Orlando, FL, USA, December 17-20, 2018}},
  \bibfield{editor}{\bibinfo{person}{M.~Arif Wani}, \bibinfo{person}{Mehmed~M.
  Kantardzic}, \bibinfo{person}{Moamar~Sayed Mouchaweh},
  \bibinfo{person}{Jo{\~{a}}o Gama}, {and} \bibinfo{person}{Edwin Lughofer}}
  (Eds.). \bibinfo{publisher}{{IEEE}}, \bibinfo{pages}{757--762}.
\newblock


\bibitem[Sennrich et~al\mbox{.}(2016)]%
        {BPE}
\bibfield{author}{\bibinfo{person}{Rico Sennrich}, \bibinfo{person}{Barry
  Haddow}, {and} \bibinfo{person}{Alexandra Birch}.}
  \bibinfo{year}{2016}\natexlab{}.
\newblock \showarticletitle{Neural Machine Translation of Rare Words with
  Subword Units}. In \bibinfo{booktitle}{\emph{Proceedings of the 54th Annual
  Meeting of the Association for Computational Linguistics, {ACL} 2016, August
  7-12, 2016, Berlin, Germany, Volume 1: Long Papers}}. \bibinfo{publisher}{The
  Association for Computer Linguistics}.
\newblock


\bibitem[Sui and Xue(2016)]%
        {SVF}
\bibfield{author}{\bibinfo{person}{Yulei Sui} {and} \bibinfo{person}{Jingling
  Xue}.} \bibinfo{year}{2016}\natexlab{}.
\newblock \showarticletitle{{SVF:} interprocedural static value-flow analysis
  in {LLVM}}. In \bibinfo{booktitle}{\emph{Proceedings of the 25th
  International Conference on Compiler Construction, {CC} 2016, Barcelona,
  Spain, March 12-18, 2016}}, \bibfield{editor}{\bibinfo{person}{Ayal Zaks}
  {and} \bibinfo{person}{Manuel~V. Hermenegildo}} (Eds.).
  \bibinfo{publisher}{{ACM}}, \bibinfo{pages}{265--266}.
\newblock


\bibitem[Tai et~al\mbox{.}(2015)]%
        {Tree-LSTM}
\bibfield{author}{\bibinfo{person}{Kai~Sheng Tai}, \bibinfo{person}{Richard
  Socher}, {and} \bibinfo{person}{Christopher~D. Manning}.}
  \bibinfo{year}{2015}\natexlab{}.
\newblock \showarticletitle{Improved Semantic Representations From
  Tree-Structured Long Short-Term Memory Networks}. In
  \bibinfo{booktitle}{\emph{Proceedings of the 53rd Annual Meeting of the
  Association for Computational Linguistics and the 7th International Joint
  Conference on Natural Language Processing of the Asian Federation of Natural
  Language Processing, {ACL} 2015, July 26-31, 2015, Beijing, China, Volume 1:
  Long Papers}}. \bibinfo{publisher}{The Association for Computer Linguistics},
  \bibinfo{pages}{1556--1566}.
\newblock


\bibitem[Tu et~al\mbox{.}(2019)]%
        {abs-1906-04881}
\bibfield{author}{\bibinfo{person}{Ming Tu}, \bibinfo{person}{Jing Huang},
  \bibinfo{person}{Xiaodong He}, {and} \bibinfo{person}{Bowen Zhou}.}
  \bibinfo{year}{2019}\natexlab{}.
\newblock \showarticletitle{Multiple instance learning with graph neural
  networks}.
\newblock \bibinfo{journal}{\emph{CoRR}}  \bibinfo{volume}{abs/1906.04881}
  (\bibinfo{year}{2019}).
\newblock
\showeprint[arXiv]{1906.04881}
\urldef\tempurl%
\url{http://arxiv.org/abs/1906.04881}
\showURL{%
\tempurl}


\bibitem[Vaswani et~al\mbox{.}(2017)]%
        {Transformer}
\bibfield{author}{\bibinfo{person}{Ashish Vaswani}, \bibinfo{person}{Noam
  Shazeer}, \bibinfo{person}{Niki Parmar}, \bibinfo{person}{Jakob Uszkoreit},
  \bibinfo{person}{Llion Jones}, \bibinfo{person}{Aidan~N. Gomez},
  \bibinfo{person}{Lukasz Kaiser}, {and} \bibinfo{person}{Illia Polosukhin}.}
  \bibinfo{year}{2017}\natexlab{}.
\newblock \showarticletitle{Attention is All you Need}. In
  \bibinfo{booktitle}{\emph{Advances in Neural Information Processing Systems
  30: Annual Conference on Neural Information Processing Systems 2017, December
  4-9, 2017, Long Beach, CA, {USA}}},
  \bibfield{editor}{\bibinfo{person}{Isabelle Guyon}, \bibinfo{person}{Ulrike
  von Luxburg}, \bibinfo{person}{Samy Bengio}, \bibinfo{person}{Hanna~M.
  Wallach}, \bibinfo{person}{Rob Fergus}, \bibinfo{person}{S.~V.~N.
  Vishwanathan}, {and} \bibinfo{person}{Roman Garnett}} (Eds.).
  \bibinfo{pages}{5998--6008}.
\newblock


\bibitem[Velickovic et~al\mbox{.}(2018)]%
        {GAT}
\bibfield{author}{\bibinfo{person}{Petar Velickovic}, \bibinfo{person}{Guillem
  Cucurull}, \bibinfo{person}{Arantxa Casanova}, \bibinfo{person}{Adriana
  Romero}, \bibinfo{person}{Pietro Li{\`{o}}}, {and} \bibinfo{person}{Yoshua
  Bengio}.} \bibinfo{year}{2018}\natexlab{}.
\newblock \showarticletitle{Graph Attention Networks}. In
  \bibinfo{booktitle}{\emph{6th International Conference on Learning
  Representations, {ICLR} 2018, Vancouver, BC, Canada, April 30 - May 3, 2018,
  Conference Track Proceedings}}. \bibinfo{publisher}{OpenReview.net}.
\newblock


\bibitem[Viega et~al\mbox{.}(2000)]%
        {ITS4}
\bibfield{author}{\bibinfo{person}{John Viega}, \bibinfo{person}{J.~T. Bloch},
  \bibinfo{person}{Y. Kohno}, {and} \bibinfo{person}{Gary McGraw}.}
  \bibinfo{year}{2000}\natexlab{}.
\newblock \showarticletitle{{ITS4:} {A} Static Vulnerability Scanner for {C}
  and {C++} Code}. In \bibinfo{booktitle}{\emph{16th Annual Computer Security
  Applications Conference {(ACSAC} 2000), 11-15 December 2000, New Orleans,
  Louisiana, {USA}}}. \bibinfo{publisher}{{IEEE} Computer Society},
  \bibinfo{pages}{257}.
\newblock


\bibitem[Wang et~al\mbox{.}(2021)]%
        {9293321}
\bibfield{author}{\bibinfo{person}{Huanting Wang}, \bibinfo{person}{Guixin Ye},
  \bibinfo{person}{Zhanyong Tang}, \bibinfo{person}{Shin~Hwei Tan},
  \bibinfo{person}{Songfang Huang}, \bibinfo{person}{Dingyi Fang},
  \bibinfo{person}{Yansong Feng}, \bibinfo{person}{Lizhong Bian}, {and}
  \bibinfo{person}{Zheng Wang}.} \bibinfo{year}{2021}\natexlab{}.
\newblock \showarticletitle{Combining Graph-Based Learning With Automated Data
  Collection for Code Vulnerability Detection}.
\newblock \bibinfo{journal}{\emph{{IEEE} Trans. Inf. Forensics Secur.}}
  \bibinfo{volume}{16} (\bibinfo{year}{2021}), \bibinfo{pages}{1943--1958}.
\newblock


\bibitem[Wheeler(2021)]%
        {Flawfinder}
\bibfield{author}{\bibinfo{person}{D.~A. Wheeler}.}
  \bibinfo{year}{2021}\natexlab{}.
\newblock \bibinfo{title}{Flawfinder}.
\newblock \bibinfo{howpublished}{\url{https://dwheeler.com/flawfinder/}, title
  = {Flawfinder.}}.
\newblock


\bibitem[Wu et~al\mbox{.}(2022)]%
        {vulcnn}
\bibfield{author}{\bibinfo{person}{Yueming Wu}, \bibinfo{person}{Deqing Zou},
  \bibinfo{person}{Shihan Dou}, \bibinfo{person}{Wei Yang},
  \bibinfo{person}{Duo Xu}, {and} \bibinfo{person}{Hai Jin}.}
  \bibinfo{year}{2022}\natexlab{}.
\newblock \showarticletitle{VulCNN: An Image-inspired Scalable Vulnerability
  Detection System}. In \bibinfo{booktitle}{\emph{44th {IEEE/ACM} 44th
  International Conference on Software Engineering, {ICSE} 2022, Pittsburgh,
  PA, USA, May 25-27, 2022}}. \bibinfo{publisher}{{ACM}},
  \bibinfo{pages}{2365--2376}.
\newblock


\bibitem[Yamaguchi et~al\mbox{.}(2014)]%
        {yamaguchi2014modeling}
\bibfield{author}{\bibinfo{person}{Fabian Yamaguchi}, \bibinfo{person}{Nico
  Golde}, \bibinfo{person}{Daniel Arp}, {and} \bibinfo{person}{Konrad Rieck}.}
  \bibinfo{year}{2014}\natexlab{}.
\newblock \showarticletitle{Modeling and Discovering Vulnerabilities with Code
  Property Graphs}. In \bibinfo{booktitle}{\emph{2014 {IEEE} Symposium on
  Security and Privacy, {SP} 2014}}. \bibinfo{publisher}{{IEEE} Computer
  Society}, \bibinfo{pages}{590--604}.
\newblock


\bibitem[Yan et~al\mbox{.}(2018)]%
        {YanWGFLH18}
\bibfield{author}{\bibinfo{person}{Yongluan Yan}, \bibinfo{person}{Xinggang
  Wang}, \bibinfo{person}{Xiaojie Guo}, \bibinfo{person}{Jiemin Fang},
  \bibinfo{person}{Wenyu Liu}, {and} \bibinfo{person}{Junzhou Huang}.}
  \bibinfo{year}{2018}\natexlab{}.
\newblock \showarticletitle{Deep Multi-instance Learning with Dynamic Pooling}.
  In \bibinfo{booktitle}{\emph{Proceedings of The 10th Asian Conference on
  Machine Learning, {ACML} 2018, Beijing, China, November 14-16, 2018}}
  \emph{(\bibinfo{series}{Proceedings of Machine Learning Research},
  Vol.~\bibinfo{volume}{95})}, \bibfield{editor}{\bibinfo{person}{Jun Zhu}
  {and} \bibinfo{person}{Ichiro Takeuchi}} (Eds.). \bibinfo{publisher}{{PMLR}},
  \bibinfo{pages}{662--677}.
\newblock
\urldef\tempurl%
\url{http://proceedings.mlr.press/v95/yan18a.html}
\showURL{%
\tempurl}


\bibitem[Ying et~al\mbox{.}(2019)]%
        {GNNExplainer}
\bibfield{author}{\bibinfo{person}{Zhitao Ying}, \bibinfo{person}{Dylan
  Bourgeois}, \bibinfo{person}{Jiaxuan You}, \bibinfo{person}{Marinka Zitnik},
  {and} \bibinfo{person}{Jure Leskovec}.} \bibinfo{year}{2019}\natexlab{}.
\newblock \showarticletitle{GNNExplainer: Generating Explanations for Graph
  Neural Networks}. In \bibinfo{booktitle}{\emph{Advances in Neural Information
  Processing Systems 32: Annual Conference on Neural Information Processing
  Systems 2019, NeurIPS 2019, December 8-14, 2019, Vancouver, BC, Canada}},
  \bibfield{editor}{\bibinfo{person}{Hanna~M. Wallach}, \bibinfo{person}{Hugo
  Larochelle}, \bibinfo{person}{Alina Beygelzimer}, \bibinfo{person}{Florence
  d'Alch{\'{e}}{-}Buc}, \bibinfo{person}{Emily~B. Fox}, {and}
  \bibinfo{person}{Roman Garnett}} (Eds.). \bibinfo{pages}{9240--9251}.
\newblock


\bibitem[Zhang et~al\mbox{.}(2008)]%
        {ZhangSPN08}
\bibfield{author}{\bibinfo{person}{Yi Zhang}, \bibinfo{person}{Arun~C.
  Surendran}, \bibinfo{person}{John~C. Platt}, {and} \bibinfo{person}{Mukund
  Narasimhan}.} \bibinfo{year}{2008}\natexlab{}.
\newblock \showarticletitle{Learning from multi-topic web documents for
  contextual advertisement}. In \bibinfo{booktitle}{\emph{Proceedings of the
  14th {ACM} {SIGKDD} International Conference on Knowledge Discovery and Data
  Mining, Las Vegas, Nevada, USA, August 24-27, 2008}},
  \bibfield{editor}{\bibinfo{person}{Ying Li}, \bibinfo{person}{Bing Liu},
  {and} \bibinfo{person}{Sunita Sarawagi}} (Eds.). \bibinfo{publisher}{{ACM}},
  \bibinfo{pages}{1051--1059}.
\newblock
\urldef\tempurl%
\url{https://doi.org/10.1145/1401890.1402015}
\showDOI{\tempurl}


\bibitem[Zheng et~al\mbox{.}(2021)]%
        {ZhengPLBEYLMS21}
\bibfield{author}{\bibinfo{person}{Yunhui Zheng}, \bibinfo{person}{Saurabh
  Pujar}, \bibinfo{person}{Burn~L. Lewis}, \bibinfo{person}{Luca Buratti},
  \bibinfo{person}{Edward~A. Epstein}, \bibinfo{person}{Bo Yang},
  \bibinfo{person}{Jim Laredo}, \bibinfo{person}{Alessandro Morari}, {and}
  \bibinfo{person}{Zhong Su}.} \bibinfo{year}{2021}\natexlab{}.
\newblock \showarticletitle{{D2A:} {A} Dataset Built for AI-Based Vulnerability
  Detection Methods Using Differential Analysis}. In
  \bibinfo{booktitle}{\emph{43rd {IEEE/ACM} International Conference on
  Software Engineering: Software Engineering in Practice, {ICSE} {(SEIP)} 2021,
  Madrid, Spain, May 25-28, 2021}}. \bibinfo{publisher}{{IEEE}},
  \bibinfo{pages}{111--120}.
\newblock
\urldef\tempurl%
\url{https://doi.org/10.1109/ICSE-SEIP52600.2021.00020}
\showDOI{\tempurl}


\bibitem[Zhou et~al\mbox{.}(2019)]%
        {devign}
\bibfield{author}{\bibinfo{person}{Yaqin Zhou}, \bibinfo{person}{Shangqing
  Liu}, \bibinfo{person}{Jing~Kai Siow}, \bibinfo{person}{Xiaoning Du}, {and}
  \bibinfo{person}{Yang Liu}.} \bibinfo{year}{2019}\natexlab{}.
\newblock \showarticletitle{Devign: Effective Vulnerability Identification by
  Learning Comprehensive Program Semantics via Graph Neural Networks}. In
  \bibinfo{booktitle}{\emph{Advances in Neural Information Processing Systems
  32: Annual Conference on Neural Information Processing Systems 2019, NeurIPS
  2019}}. \bibinfo{pages}{10197--10207}.
\newblock


\end{thebibliography}

\appendix

\end{document}